# Breakdown of the Kirchhoff's law of thermal radiation by a spatiotemporally modulated nonreciprocal metasurface

Anatoly Efimov[1], Chun-Chieh Chang[1], Simo Pajovic[2,3], Wilton J. M. Kort-Kamp[2], Dongsung Kim[4], Hou-Tong Chen[1], Diego A. R. Dalvit[2]*, & Abul K. Azad[1]*

[1]Center for Integrated Nanotechnologies, Los Alamos National Laboratory, Los Alamos, NM, 87545, USA

[2]Theoretical Division, Los Alamos National Laboratory, Los Alamos, NM, 87545, USA

[3]Department of Mechanical Engineering, Massachusetts Institute of Technology, Cambridge, MA, 02138, USA

[4]Accelerator Operations and Technology Division, Los Alamos National Laboratory, Los Alamos, NM, 87545, USA

*Corresponding authors: aazad@lanl.gov, dalvit@lanl.gov

**Abstract**

**Kirchhoff's law of thermal radiation, which dictates that the emissivity of a surface equals its absorptivity under thermal equilibrium, fundamentally limits the efficiency of photonic systems by enforcing reciprocal energy exchange between source and detector. Breaking this reciprocity is important for advancing photonic devices for energy conversion, radiative cooling, and mid-infrared sensing and imaging. Driven by the growing need for photonic platforms to overcome reciprocity constraints, we report the first demonstration of spatiotemporally modulated nonreciprocal metasurfaces operating at mid-infrared frequencies enabling the violation of the Kirchhoff's law at room temperature. We fabricate a graphene-based integrated photonic structure and experimentally demonstrate nonreciprocal reflection from a metasurface modulated at gigahertz frequencies. We further develop a theoretical framework to relate nonreciprocal scattering under spatiotemporal modulation with unequal absorptivity and emissivity for violation of the spectral directional Kirchhoff's law. Together, our experiment and theory imply effective decoupling of absorption and emission channels by breaking time-reversal symmetry at thermal wavelengths, thereby representing an indirect demonstration of breakdown of Kirchhoff's law of thermal radiation.**

**Introduction**

In electromagnetism, the Lorentz reciprocity theorem states that a source and a detector of light can swap positions without changing the outcome of the measurement by the detector[1-3]. In other words, the scattering matrix is symmetric, $S_{ab} = S_{ba}$. Reciprocity assumes that light propagates in a linear, time-invariant medium with symmetric permittivity, permeability, and conductivity tensors. Most devices operate under the regime of reciprocity, which can be useful since it implies symmetry between emission and absorption—in fact, reciprocity underlies Kirchhoff's law of thermal radiation (henceforth referred to as Kirchhoff's law), which states that the spectral directional emissivity and absorptivity of a surface are equal[4,5]: $e(\omega, \theta, \phi) = a(\omega, \theta, \phi)$, where $e$ is emissivity, $a$ is absorptivity, $\omega$ is (angular) frequency, and $\theta$ and $\phi$ are polar and azimuthal directions, respectively. However, reciprocity can have undesirable effects as well. Examples include solar cells re-emitting absorbed solar energy, radiative coolers absorbing thermal radiation, and antennas hearing their own echoes.

In principle, a nonreciprocal system for which $S_{ab} \neq S_{ba}$, achieved by lifting one or more of the assumptions of the Lorentz reciprocity theorem[1,2], could circumvent these issues. In the past, this has been realized using magneto-optic materials, which break time reversal symmetry and have antisymmetric permittivity tensors. However, the need for external bias using bulk magnets can be cumbersome and limit their applicability to integrated systems[6-13]. Nonlinear materials have achieved some success, but these systems are constrained by significant power requirements and long interaction lengths[14-16]. Spatiotemporal modulation has been one of the most successful approaches to nonreciprocity (Fig. 1a), having been theoretically predicted and even experimentally demonstrated in waveguides[17-19], antennas[20,21], and metasurfaces[22]. Among these systems, spatiotemporally modulated metasurfaces (STMMs) are particularly attractive due to their integrability and size, weight, and power advantages compared to magneto-optic and nonlinear materials. In principle, STMMs offer complete control over scattering amplitude, phase, frequency, direction, and polarization of light, all in a lightweight, ultrathin platform. More importantly, the optical properties of STMMs can be actively or dynamically tuned over time and/or locally to continuously adapt to their surroundings.

Nonreciprocity has been theoretically and experimentally explored in the microwave and far-infrared spectra but extending it to the mid-infrared (mid-IR) spectrum has significant implications for radiative heat transfer. Fundamentally, nonreciprocal systems should violate Kirchhoff's law[9,23], leading to unequal spectral directional emissivity and spectral directional absorptivity, $e(\omega, \theta, \phi) \neq a(\omega, \theta, \phi)$, as illustrated in Fig. 1b. This is typically shown by considering the energy balance between a graybody and a blackbody enclosure and demonstrating that the difference between $e(\omega, \theta, \phi)$ and $a(\omega, \theta, \phi)$ is nonzero. For planar magneto-optic materials, this equals the difference in reflectivities in opposite propagation directions, $\rho(\omega, \theta, \phi + \pi) - \rho(\omega, \theta, \phi)$[23-26]. However, this relation may not be true in general and has not been extended to STMMs. A corollary of the violation of Kirchhoff's law is directionally asymmetric emission and absorption in planar systems, i.e., $e(\omega, \theta, \phi) \neq e(\omega, \theta, \phi + \pi)$ and $a(\omega, \theta, \phi) \neq a(\omega, \theta, \phi + \pi)$. This implies that nonreciprocity can enable highly directional or even unidirectional heat flow[27-29]. This has the potential to transform technologies such as solar energy harvesting, which can approach the thermodynamic limit[30] in nonreciprocal systems[31,32], active and dynamic thermal management, radiative cooling and optical refrigeration[33,34], and bioinspired, adaptive thermal camouflage.

Despite the promise of mid-infrared nonreciprocity and its relevance to thermal radiation, there have been few experiments demonstrating nonreciprocal emission or absorption in this spectral range. Almost all have used magneto-optic materials and large magnetic fields on the order of 1 T (comparable to an MRI scanner). To date, the violation of Kirchhoff's law has been directly demonstrated in experiments by measuring $e(\omega, \theta, \phi) - a(\omega, \theta, \phi)$ in magneto-optic materials only a handful of times[9,10,12,13], although there have been numerous indirect demonstrations through measurements of $\rho(\omega, \theta, \phi + \pi) - \rho(\omega, \theta, \phi)$ [8,11,24,35]. Recently, a direct measurement of the breakdown of Kirchhoff's law has been reported using a nonlinear GaAs crystal[36]. Even though nonreciprocal beam steering[37,38] and nonreciprocal (i.e., directionally asymmetric) reflection[22] have been achieved in the microwave spectrum via spatiotemporal modulation, mid-infrared nonreciprocity has not been realized, primarily because of the challenging requirements of modulation frequencies on the order of 1–10 GHz[39,40] and sub-10 μm physical dimensions commensurate with mid-infrared wavelengths. Few materials can meet both requirements, and the task of integrating spatiotemporal modulation, i.e., applying an external bias, into such small structures (which may be challenging to fabricate in the first place) is highly nontrivial.

In this work, we experimentally demonstrate nonreciprocal frequency conversion using an STMM designed for operation at a wavelength of 10 μm, near the peak wavelength of room-temperature thermal radiation. By setting up our STMM in the Littrow configuration[41], we can measure the amplitude, frequency, and propagation direction of synthetically diffracted modes. Using this approach, we show that synthetic diffraction produces unidirectional frequency up- and down-conversion, i.e., from $\omega$ to $\omega \pm \Omega$ for the first-order mode, where $\Omega/2\pi = 1$ GHz is the modulation frequency. This is the first experimental demonstration of synthetic diffraction in the mid-infrared spectrum and at gigahertz modulation frequency as compared to the MHz modulation frequencies used in STMMs similar to ours[22,42]. Then, we show that when the propagation direction of the first-order mode is reversed, the frequency of the incident light along the initial Littrow direction is further converted from $\omega \pm \Omega$ to $\omega \pm 2\Omega$ instead of converted back to $\omega$ along the initial direction of incidence. This is direct evidence of nonreciprocity since it means the scattering matrix is asymmetric. Finally, we discuss the implications of our work for thermal radiation and prove that nonreciprocal frequency conversion upon reflection from a spatiotemporally modulated metasurface is tantamount to violating the spectral directional Kirchhoff's law of radiation. We show that the difference between $e(\omega, \theta, \phi)$ and $a(\omega, \theta, \phi)$ is related to the difference between forward and backward scattering, summed over all possible mode conversions. However, we argue that only some terms in the summation needs to be nonzero to violate Kirchhoff's law and that an STMM with the same phase profile as our experiments would suffice, which we demonstrate using numerical simulations.

## Results

### Experimental setup

**Sample:** Our STMM consists of an array of 36 rectangular pixels of size 5.7 μm × 200 μm. Each pixel contains six metallic patch antennas placed on top of a graphene monolayer, which is transferred onto a dielectric stack of alumina and amorphous-germanium (a-Ge) that is backed by an optically thick metal ground plane (see Fig. 1c, with fabrication details in Supplementary Note S1). The sample thus represents a metal-dielectric-metal metasurface cavity working in reflection mode[43,44], with parameters optimized via electromagnetic simulations using COMSOL Multiphysics (see Supplementary Note S2). Monolayer graphene is used to modulate the antennas' optical response via electrically controlled charge density at 1 GHz frequency. Within each pixel, the rectangular metallic patches serve dual roles: they act as antennas that couple infrared light into the metasurface cavity and simultaneously function as the top electrode for modulation of the charge carrier density in graphene not covered by metal (i.e., between the patch antennas). The a-Ge spacer, which is a dielectric layer for IR but electrically conducting at 1 GHz, provides a common ground electrode for all pixels. There is a 20 nm alumina isolation layer between graphene and a-Ge, which is thin enough to allow substantial Fermi energy modulation in graphene with only a few-volt applied bias, thereby modulating the resonant reflection of the metasurface (or the scattering of each pixel)[44].

The overall resonant response of the metasurface is defined by both the patch antennas and the alumina and a-Ge layers, which is parametrically optimized in COMSOL simulations. We maximize the synthetic diffraction efficiency of the STMM (see Supplementary Note S2), achieving its peak efficiency with antenna width 850 nm, period 950 nm (hence, gap between antennas of 100 nm), and a-Ge thickness 500 nm. The latter is deeply subwavelength for both air and a-Ge, resulting in the complete absence of static diffraction orders neither above nor below the antenna layer. Although the graphene layer is pixelated to isolate individual pixels, this creates no observable static diffraction orders from the device. It is important to emphasize that all device layers carry both optical and radio frequency (RF) functionality, which necessitated extensive modeling and simulation in these two domains under realistic fabrication constraints. The sample is fabricated using a combination of film deposition, photolithography, e-beam lithography, metallization, lift-off, and reactive ion-etching, with scanning electron microscopy (SEM) images shown in Fig. 1d,e. Finally, the fabricated chip is attached to a board using the flip-chip bonding method, and all pixels are connected to the modulation circuitry (Fig. 1f).

**RF modulation:** Graphene modulation and RF pixel driving is implemented as a synthetic unidirectional traveling surface wave, which imparts its momentum and frequency onto the diffracted optical waves by upshifting (downshifting) positive (negative) synthetic diffraction orders, depending on the propagation direction of the synthetic grating along the surface. Our metasurface is designed to generate synthetic diffraction orders in reflection around a center wavelength of about 10 μm (i.e., 30 THz) when driven with properly phase-controlled GHz RF voltage signals applied to individual pixels in a 3-pixel periodic pattern. Such a periodic pattern is the simplest format allowing for directional propagation of the synthetic diffraction grating (see Supplementary Note S3), unlike 2-pixel periodic modulation, which does not offer directionality, or periodicity with 4 or more pixels, which increases experimental complexity. The voltage applied to pixel $i$ at the spatial coordinate $r_i$ is a time-harmonic function of the form $V(r_i, t) = V_0 + \Delta V \cos(\Omega t \pm \beta \cdot r_i)$, where $V_0$ is the baseline DC voltage, $\Delta V$ is the voltage modulation amplitude, $\Omega$ is the modulation frequency, $\beta$ is the spatial modulation wavevector, and the sign determines the propagation direction of the surface modulation. The magnitude of the spatial modulation vector $\beta$ is defined by the interpixel spacing $d = r_{i+1} - r_i$ and the 120° phase difference between pixels as $\beta d = 2\pi/3$.

The experiments are performed using a piezo tunable mode-hop-free quantum cascade laser (QCL, Sacher Lasertechnik). The laser is characterized by a stable single optical frequency operation with the option of tuning this frequency in a range of about 10 GHz without any mode hopping with careful operation. The first order diffracted optical beam arising from the STMM is measured in the Littrow configuration, in which the incident and detected optical beams are collinear as shown in Fig. 2. The use of the Littrow configuration provides a key simplification in a nonreciprocity experiment: in general, to demonstrate that an optical system is nonreciprocal, one has to perform a "forward" experiment sending light from input to output; then a "reverse" experiment taking the output of the forward experiment and using it as the new input for the reverse experiment, to finally demonstrate that the scattering matrix is asymmetric, $S_{ab} \neq S_{ba}$. In principle, one should swap the positions of the pump laser and detector in Fig. 2 for forward and reverse experiments. However, in the Littrow configuration the input and output optical modes are collinear, and therefore we can keep the laser and detector in fixed positions and tune the laser frequency and Fabry-Pérot etalon (FPE) filter appropriately for forward and reverse experiments.

When the pixels are driven in a 3-pixel periodic pattern, pairs of diffraction orders $m$ are produced at Littrow angles $\theta_{L,m} = \arcsin(m\lambda/2R)$, where $R = 3 \times 5.7$ μm $= 17.1$ μm. We call these "synthetic" orders. Our STMM generates $m = \pm 1$ synthetic diffraction orders at Littrow angles $\theta_{L,m=\pm1} = \pm 16.1°$ at a wavelength $\lambda = 10$ μm. Importantly, with pixels dynamically driven at 120° RF phase relative to their neighbors, the synthetic orders carry optical signals with their optical frequencies up- and down-shifted, respectively. The direction of the frequency shift (up or down by the 1 GHz RF modulation) depends on RF phase assignment to the three pixels in each period, i.e., $\cdots [-120°, 0, +120°], \cdots$ or $\cdots, [+120°, 0, -120°], \cdots$. We can easily switch between these two configurations via computer control.

**Synthetic diffraction and frequency conversion in the mid-IR spectrum**

Since there is no simple way to measure the absolute optical frequencies of the synthetic orders to a 1 GHz precision, we employ an FPE filter with a free spectral range (FSR) of 3 GHz. The passband is tuned by physically rotating the filter in the plane containing the optical beam. The transmission of the FPE displays a series of peaks as a function of the FPE angle with respect to the optical axis. The synthetically diffracted laser beam is propagated through the FPE and is focused on a liquid nitrogen (LN2) cooled mercury cadmium telluride (MCT) detector. The detector output is connected to a lock-in amplifier referenced to a 10 kHz frequency, which is also used to modulate the 1 GHz RF driving field applied to the pixels. We detect the difference between the two levels of the synthetic signal: 1) during the first 50 μs no RF is applied, corresponding to no synthetic signal present, and 2) during the second 50 μs RF is applied, corresponding to synthetic diffraction present. The lock-in amplifier effectively measures the difference between these two levels of detector output. These measurements are conducted while we tune both the laser frequency and the FPE angle in sequence, which results in the 2D density maps shown in Fig. 3. The horizontal axis in these maps corresponds to the frequency of the input QCL laser and the vertical axis is the FPE angular position.

First, we perform the measurements with no RF modulation applied. To still enable Littrow configuration measurements with lock-in amplifier, a 10 kHz drive voltage, which otherwise modulates the RF signal, is directly applied to every third pixel of the metadevice to create a quasi-static grating for diffraction. The results of these quasi-static measurements are shown in Fig. 3a,d with effectively no frequency conversion—the output frequency is equal to the input frequency to within 10 kHz. This 2D density map serves as the calibration between the FPE angular position and transmitted frequency. The bright rings correspond to the transmission of a specific optical frequency at certain angles of the FPE. The map is symmetric around the central horizontal line because the FPE transmission is the same when it is tilted by the same angle, either positive or negative. The transmission is higher near 0° and lower at the periphery, which is a feature of the FPE. The multiple rings and the periodicity along the horizontal axis reflect the 3 GHz FSR of the FPE and was used for frequency-voltage tuning calibration of the QCL laser. The measured experimental efficiency of the first order diffraction is $\approx 8 \times 10^{-4}$.

We then apply a spatiotemporal modulation ($\Omega/2\pi = 1$ GHz, $\Delta V = 2.5$ V) to the STMM in the 3-pixel periodic sequence $[+120°, 0, -120°]$ and obtain the 2D map in Fig. 3b. The rings are displaced to the right exactly by $\Omega/2\pi = 1$ GHz modulation frequency, demonstrating unidirectional and essentially nonreciprocal frequency conversion. To better understand the data, we first consider the "forward" experiment by arbitrarily choosing an input frequency of the QCL, say $\omega_{in} = \omega_0 + \Delta\omega$, where $\omega_0/2\pi \approx 31.6$ THz (wavelength 9.5 μm) and $\Delta\omega/2\pi = +0.5$ GHz. A vertical slice (solid blue line) of the map in Fig. 3b at $\omega_{in}$ cuts through one of the rings at two symmetric points (with one marked with a blue circle) located at angular positions of about $\pm 0.8°$. Following the horizontal blue dashed line, in the reference map (Fig. 3a) we find that the FPE transmission frequency is $\omega_{out}/2\pi = \omega_0/2\pi - 0.5$ GHz at these angular positions, as indicated by the yellow circle and vertical dashed line. By comparing the maps in Figs. 3a and 3b, we find that this data analysis is applicable to arbitrary values of $\Delta\omega$. Hence, we conclude that frequency down-conversion $\omega_{out} = \omega_{in} - \Omega$ occurred in the spatiotemporally modulated metadevice. It is worth mentioning that there are no observable intermediate rings in Fig. 3b, suggesting a dominating $\omega_{out} = \omega_{in} - \Omega$ down-conversion, with other frequency mixing components negligible along the original Littrow direction.

The unidirectional frequency down-conversion is sufficient to prove that the metadevice supports nonreciprocity, but to be clearer we consider the "reverse" experiment while keeping the modulation protocol $[+120°, 0, -120°]$ unchanged. The Littrow configuration with colinear input and output optical modes allows us to keep the QCL laser and detector in the same positions in the reverse experiment. This leads to the same 2D optical density map depicted in

Fig. 3b. We consider the previously down-converted frequency to be the new input frequency $\omega'_{in}/2\pi = \omega_{out}/2\pi = \omega_0/2\pi - 0.5$ GHz and extend a vertical slice (yellow solid line) that intersects a ring in Fig. 3b at FPE angular positions about $\pm 1.25°$ (yellow circle for the negative one). Following the horizontal yellow dashed line to the reference map in Fig. 3a we find the intersection point marked by the red circle, which corresponds to the new output frequency $\omega'_{out}/2\pi = \omega_0/2\pi - 1.5$ GHz, suggesting yet another down-conversion, $\omega'_{out} = \omega_{in} - 2\Omega$. Importantly, there is no intersection between the horizontal yellow dashed line with any of the FPE transmission rings in Fig. 3a at $\omega_0/2\pi + 0.5$ GHz, which would have indicated an *up-conversion* back to the original input frequency of $\omega_{in}/2\pi = \omega_0/2\pi + 0.5$ GHz precisely at the Littrow direction. Figure 3c schematically illustrates the two steps of the frequency down-conversion just discussed. This key observation demonstrates nonreciprocal reflection from our graphene-based spatiotemporally modulated metasurface operating at mid-infrared frequencies.

For completeness, we have also demonstrated nonreciprocal reflection in a reversed modulation sequence $[-120°, 0, +120°]$, with the 2D optical density map shown in Fig. 3e. In this case the rings are displaced to the left by the 1 GHz modulation. Following a similar procedure as above, one can show that in the forward experiment there is frequency up-conversion $\omega_{out} = \omega_{in} + \Omega$. The reverse experiment gives another up-conversion, $\omega'_{out} = \omega_{in} + 2\Omega$ and no signal at the initial input frequency is observed. Figure 3f depicts the two steps of frequency up-conversion. As before, this indicates nonreciprocal reflection from our spatiotemporally modulated metasurface. The measured efficiency of mode conversion $\omega_0 \rightarrow \omega_{\pm 1}$ is $\approx 4 \times 10^{-5}$.

To corroborate our experimental findings, we performed full-wave COMSOL simulations of the graphene-based STMM under spatiotemporal modulation. The simulations reproduce the nonreciprocal reflection process by explicitly resolving the coupling between Floquet harmonics, showing excellent qualitative agreement with the measured forward and reverse spectra (the modeling approach is described in Supplementary Note S4). Indeed, in the forward case with a 3-pixel phase modulation $[+120°, 0, -120°]$, an incident light field of frequency $\omega_0$ impinging at the Littrow angle $\theta_{in} = \theta_L = 16.1°$ (Fig. 4a) is reflected into three main channels: specular reflection at $\omega_0$ and $\theta_{m=0} = -16.1°$ (Fig. 4b), a down-converted mode at $\omega_0 - \Omega$ and $\theta_{m=-1} = 16.1°$ (Fig. 4c), and up-converted order at $\omega_0 + \Omega$ and $\theta_{m=+1} = -56.3°$ (Fig. 4d). As expected, only the down-converted field is colinear with the incident field. In the reverse scenario, the down-converted beam illuminates the metasurface at the Littrow angle (Fig. 4e), generating specular reflection at $\omega_0 - \Omega$ with an angle of $-16.1°$ (Fig. 4f), a further down-converted field at $\omega_0 - 2\Omega$ and angle of $16.1°$ (Fig. 4g), and an up-converted mode of frequency $\omega_0$ scattered in the $-56.3°$ direction (Fig. 4h). These results clearly demonstrate unidirectional frequency conversion and breaking of reciprocity.

**Implications of spatiotemporal modulation for thermal radiation**

Given that our STMM was designed to control light in the mid-IR spectrum, we discuss the implications of our work for thermal radiation, particularly Kirchhoff's law. We prove that nonreciprocal frequency conversion upon reflection (synthetic diffraction) from an STMM is equivalent to the violation of Kirchhoff's law by extending derivations found in the literature for static problems[23-26]. Here, we provide an outline of the proof; more detailed steps can be found in Supplementary Note S5. Consider a small, opaque graybody which is spatiotemporally modulated and surrounded by a unit hemispherical enclosure which is black for frequencies $\omega \in [\omega_m, \omega_m + d\omega]$ and perfectly reflecting otherwise. Here, $\omega_m = \omega_0 + m\Omega$ ($m \in \mathbb{Z}$) are frequency harmonics of the modulation frequency. The system, showing in Fig. 4i, is at thermodynamic equilibrium. Essentially, the emissivity of the enclosure is a frequency comb such that it emits and absorbs only at those frequencies which the graybody can interact with because of spatiotemporal modulation. An incoming plane wave of frequency $\omega_0$ and in-plane wavevector $\boldsymbol{k}_{\parallel,0}$ reflects into different frequency harmonics and associated diffraction orders with in-plane wavevector $\boldsymbol{k}_{\parallel,m} = \boldsymbol{k}_{\parallel,0} + m\boldsymbol{\beta}$. In addition, each wavevector $\boldsymbol{k}_m = \boldsymbol{k}_{\parallel,m} + k_{z,m}\hat{\boldsymbol{z}}$ (where $k_{z,m} = \sqrt{(\omega_m/c) - |\boldsymbol{k}_{\parallel,m}|^2}$) has corresponding solid angles $d\Omega_{dA \to dA_m} = dA_m \cos\theta_m$ and $d\Omega_{dA_m \to dA} = dA \cos\theta_m$. This results in a "polka dot pattern" of differential areas $dA_m$ which emit and receive light on the enclosure. The graybody emits, absorbs, and reflects light along direction vectors $\hat{\boldsymbol{n}}_m = \sin\theta_m \cos\phi_m \hat{\boldsymbol{x}} + \sin\theta_m \sin\phi_m \hat{\boldsymbol{y}} + \cos\theta_m \hat{\boldsymbol{z}}$, where $\theta_m$ and $\phi_m$ are polar and azimuthal angles of incidence associated with spectral-directional channel $m$. We define the reflected direction vector as well, in which the sign of the z-component is flipped: $\hat{\boldsymbol{n}}'_m = \sin\theta_m \cos\phi_m \hat{\boldsymbol{x}} + \sin\theta_m \sin\phi_m \hat{\boldsymbol{y}} - \cos\theta_m \hat{\boldsymbol{z}}$ (see Fig. S5.2 in the Supplementary Information).

We are interested in establishing a relationship between the light emitted and absorbed by the graybody (in other words, establish a generalized Kirchhoff's law) and show that it is not equality. In the perturbative regime of small modulation frequency and amplitude, there is no energy exchange between the source of the modulation and the STMM and/or the electromagnetic field. In this approximation, the source simply tunes the instantaneous optical properties of the metasurface device. Consider the light that leaves the graybody toward the set of all possible differential areas $dA_m$ on the enclosure, or "receiver polka dots": this includes emitted light from $dA$ and reflected light from all possible $dA_m$'s on the enclosure, or "emitter polka dots." In general, the radiant power of the reflected light can be written as

$$\sum_m \left[ \sum_n \rho(\omega_n \to \omega_m, \widehat{\boldsymbol{n}}_n \to \widehat{\boldsymbol{n}}'_m) \, I_b(\omega_n, T) \, dA \, dA_n \cos\theta_n \right] dA_m \cos\theta_m \quad (1)$$

where $\rho(\omega_n \to \omega_m, \widehat{\boldsymbol{n}}_n \to \widehat{\boldsymbol{n}}'_m)$ is the bidirectional reflectance distribution function, defined on the basis of both incoming and outgoing frequency. This, plus the radiant power of the light emitted by the graybody, must equal that of the light emitted by all possible $dA_m$'s because of thermodynamic equilibrium, resulting in the equation

$$\sum_m I_b(\omega_m, T) \, dA \, dA_m \cos\theta_m = \sum_m e(\omega_m, \widehat{\boldsymbol{n}}'_m) \, I_b(\omega_m, T) \, dA \, dA_m \cos\theta_m$$
$$+ \sum_m \left[ \sum_n \rho(\omega_n \to \omega_m, \widehat{\boldsymbol{n}}_n \to \widehat{\boldsymbol{n}}'_m) \, I_b(\omega_n, T) \, dA \, dA_n \cos\theta_n \right] dA_m \cos\theta_m, \quad (2)$$

where $I_b(\omega_m, T)$ is the blackbody spectral radiance and $e(\omega_m, \widehat{\boldsymbol{n}}'_m)$ is the spectral directional emissivity. By rearranging Eq. (2) and arguing that $e(\omega_m, \widehat{\boldsymbol{n}}'_m) + \sum_n \rho(\omega_n \to \omega_m, \widehat{\boldsymbol{n}}_n \to \widehat{\boldsymbol{n}}'_m) \frac{I_b(\omega_n, T)}{I_b(\omega_m, T)} dA_n \cos\theta_n \leq 1$ (otherwise, the enclosure is receiving more light than it can possibly emit), it can be shown that

$$0 = 1 - e(\omega_m, \widehat{\boldsymbol{n}}'_m) - \sum_n \rho(\omega_n \to \omega_m, \widehat{\boldsymbol{n}}_n \to \widehat{\boldsymbol{n}}'_m) \frac{I_b(\omega_n, T)}{I_b(\omega_m, T)} dA_n \cos\theta_n. \quad (3)$$

Similarly, consider the light that leaves $dA_m$ and arrives at $dA$. It includes light emitted by all possible $dA_m$'s that is absorbed and reflected by the graybody, which, after some manipulation, gives us a second equation:

$$0 = 1 - a(\omega_m, -\widehat{\boldsymbol{n}}'_m) - \sum_n \rho(\omega_m \to \omega_n, -\widehat{\boldsymbol{n}}'_m \to -\widehat{\boldsymbol{n}}_n) \, dA_n \cos\theta_n, \quad (4)$$

where $a(\omega_m, -\widehat{\boldsymbol{n}}'_m)$ is the spectral directional absorptivity. Subtracting Eq. (3) from Eq. (4), and assuming $I_b(\omega_n, T) \approx I_b(\omega_m, T)$ and $m - n \ll \Omega/\omega_0$, we arrive at

$$e(\omega_m, \widehat{\boldsymbol{n}}'_m) - a(\omega_m, -\widehat{\boldsymbol{n}}'_m)$$

$$= \sum_n [\rho(\omega_m \to \omega_n, -\widehat{\boldsymbol{n}}'_m \to -\widehat{\boldsymbol{n}}_n) - \rho(\omega_n \to \omega_m, \widehat{\boldsymbol{n}}_n \to \widehat{\boldsymbol{n}}'_m)] \, dA_n \cos \theta_n, \quad (5)$$

valid for any spatiotemporal mode $m$. Equation (5) is the key result of our theory and represents a generalized Kirchhoff's law of thermal radiation for weakly spatiotemporally modulated systems. It has a relatively simple interpretation: in spatiotemporally modulated systems, for each mode $m$, $e(\omega_m, \widehat{\boldsymbol{n}}'_m) \neq a(\omega_m, -\widehat{\boldsymbol{n}}'_m)$ (meaning the spectral directional Kirchhoff's law is violated) and their difference is equal to the energy that is nonreciprocally scattered into the spectral directional channels created by spatiotemporal modulation. This proves that frequency or mode conversion upon reflection is equivalent to the violation of Kirchhoff's law, since it is no longer true in this case that $e(\omega_m, \widehat{\boldsymbol{n}}'_m) = a(\omega_m, -\widehat{\boldsymbol{n}}'_m)$.

The breakdown of the spectral directional Kirchhoff's law (a nonzero value of the left-hand side of Eq. (5)) occurs if and only if the right-hand side is nonzero. For a generic spatiotemporally modulated metasurface, this requires the sum over a large number of scattering modes of the difference between forward and reverse scattering (spectral directional nonreciprocity) to vanish. This poses an important experimental challenge because it requires measuring nonreciprocal reflection over various spatiotemporal scattering modes. Fig. 4j shows the numerical evaluation of the differences $\Delta \rho_{mn} = \rho(\omega_m \to \omega_n, -\widehat{\boldsymbol{n}}'_m \to -\widehat{\boldsymbol{n}}_n) - \rho(\omega_n \to \omega_m, \widehat{\boldsymbol{n}}_n \to \widehat{\boldsymbol{n}}'_m)$ for our experimental travelling-wave modulation. We consider input mode $m = 0$ with frequency $\omega_0/2\pi = 30$ THz at the Littrow direction of incidence $-\widehat{\boldsymbol{n}}'_0$, corresponding to polar angle of incidence $\theta_{\text{in}} = \theta_L = 16.1°$. The term $n = m = 0$ corresponds to no frequency conversion and specular reflection, having identical forward and reverse reflectivities and thus no net contribution to the violation of Kirchhoff's law via Eq. (5). The terms $n = \pm 1$ give $\rho(\omega_0 \to \omega_{\pm 1}, -\hat{\boldsymbol{n}}_0' \to -\hat{\boldsymbol{n}}_{\pm 1}) \approx 3 \times 10^{-5}$ and $\rho(\omega_{\pm 1} \to \omega_0, \hat{\boldsymbol{n}}_{\pm 1} \to \hat{\boldsymbol{n}}_0') = 0$, providing the major contribution to the violation of Kirchoff's law. All other mode conversions with $|n| \geq 2$ give subleading contributions. Our experimental demonstration of Lorentz nonreciprocity in reflection together with our developed theory and numerics are tantamount to an indirect demonstration of the violation of Kirchhoff's law of thermal radiation at room temperature.

**Discussion**

In summary, we have introduced a graphene-based STMM platform modulated at GHz frequencies for experimental demonstration of breakdown of Lorentz reciprocity at thermal mid-IR wavelengths. We have also developed a theory that relates nonreciprocal reflection with the breakdown of the spectral directional Kirchhoff's law of thermal radiation. Jointly, our experiment and theory represent an indirect demonstration of effective decoupling of absorption and emission channels by breaking time-reversal symmetry at thermal wavelengths. Our custom-designed driver electronics and STMM device architecture have enabled high modulation speeds and the contingent requirements of physical dimensions commensurate with mid-IR wavelengths. Our work has the potential to find applications in mid-IR optical isolators, solar energy harvesting, thermophotovoltaics for waste heat recovery, active/dynamic thermal management for electronics, radiative cooling, and adaptive thermal camouflage and thermal signatures.

Unlike magneto-optic materials, STMMs enable "total" nonreciprocity in the sense that scattering changes the very nature of the electromagnetic modes in opposite propagation directions, e.g., photon-to-photon conversion in one direction and photon-to-surface-wave conversion in the other[22]. Furthermore, our extension of previous derivations of the violation of

Kirchhoff's law to STMMs that properly accounts for conversion between modes (or lack thereof) is significant as it offers insights into the relationship between emission and absorption in spatiotemporally modulated systems and provides an avenue for indirect demonstrations of the violation of Kirchhoff's law. Our generalized Kirchhoff's law for nonreciprocal STMMs also facilitates thermal photonic design by circumventing direct numerical simulations of thermal radiation, which are computationally expensive[45,46].

Finally, we note that a direct demonstration of the breakdown of Kirchhoff's law of thermal radiation at mid-IR frequencies will ultimately require independent measurements of absorptivity and emissivity. While such measurements are challenging with our present sample largely due to limited effective modulation depth and conversion efficiency arising from the RC time constant and RF-impedance mismatching. These issues can be addressed by further improving the device design to reduce the RC time constant[44] and driver electronics for better impedance matching. Thus, we view this as an exciting opportunity for future work toward a direct demonstration of the breakdown of Kirchhoff's law and its applications in thermal management.

**Data availability**

The data that support the plots within this paper and other findings of this study are available from the corresponding authors upon reasonable request.

**Acknowledgments**

We acknowledge discussions with A. Alú, S. Boriskina, A. Dattelbaum, S. Mann, and R. Yu. This work has been supported by Los Alamos National Laboratory LDRD and other programs. Part of this work was performed at the Center for Integrated Nanotechnologies, a U.S. Department of Energy, Office of Basic Energy Sciences user facility. LANL is operated by Triad National Security, LLC, for the National Nuclear Security Administration of the U.S. Department of Energy (Contract No. 89233218CNA000001). S.P. was supported by the U.S. Department of Energy, Office of Science, Office of Workforce Development for Teachers and Scientists, Office of Science Graduate Student Research (SCGSR) program. The SCGSR program is administered by the Oak Ridge Institute for Science and Education for the DOE under contract number DE-SC0014664.

**Author contributions**

A.A. and D.D. conceived the project. A.A. and A.E. designed the sample. A.E. developed the electronics and carried out all measurements. C.C.C. and D.K. fabricated samples. D.D. and S.P. developed the theory. W.K.K. performed numerical simulations. H.T.C. and A.E. analyzed the modulation scheme. All authors contributed to the writing of the manuscript.

**Competing interests**

The authors declare no competing interests.

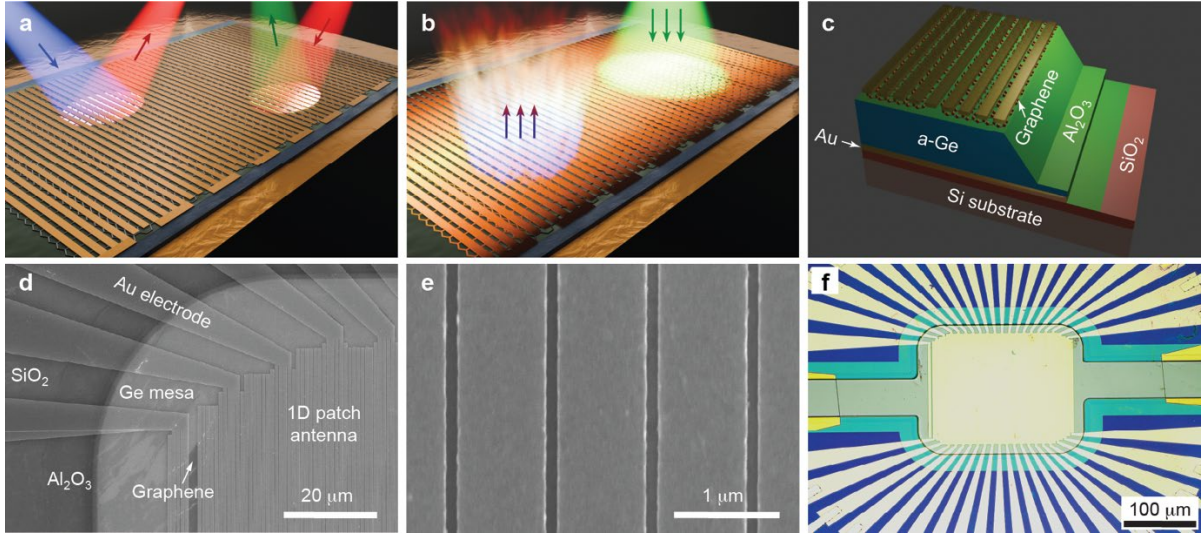

**Fig. 1 | Spatiotemporally modulated nonreciprocal metasurface for the breakdown of Kirchhoff's law of thermal radiation. a** Nonreciprocal reflection from a graphene STMM. An incident beam (blue) impinging on an STMM is downshifted and reflected into a diffraction order (red) by the metasurface. In the reverse, the red beam is not scattered back into the original blue beam, but is further down-shifted and diffracted into a new direction (green). **b,** Breakdown of Kirchhoff's law of thermal radiation. As an example, the STMM may absorb green light but only emits blue/red. **c,** Schematic of layer structure of an STTM. **d,** Top-view SEM image of a fully fabricated STMM. **e,** Top-view SEM image of one-dimensional gold patch antennas. **f,** Optical image of our STMM device.

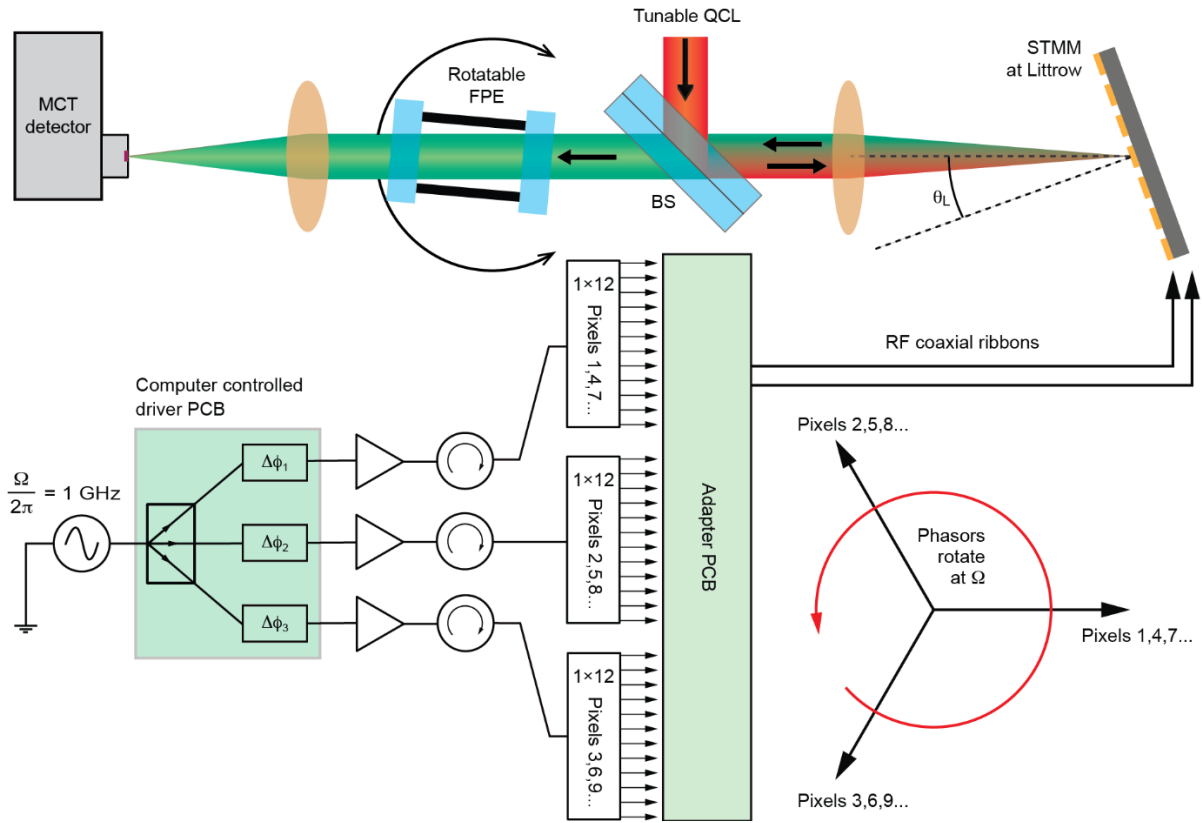

**Fig. 2 | Schematic of the experiment with angle-tuned Fabry-Pérot etalon (FPE) spectral filter in the detection path.** The STMM device is probed at Littrow angle $\theta_L$ with a tunable QCL. A 50/50 beamsplitter (BS) is used to separate incident and diffracted beams, which are collinear at Littrow configuration used here. A 1 GHz RF sinusoidal signal from a frequency synthesizer is split three ways and passed through three independent computer-controlled RF phase shifters, amplifiers, isolators and sent to three $1 \times 12$ RF power dividers. The 36 outputs from the dividers are routed to individual pixels on the STMM using a custom adapter PCB and two micro-coax ribbon cables. Bottom-right inset shows RF voltage phasors applied to the three pixels in each spatial period. Direction of phasor rotation is defined by the modulation protocol being either $[-120°, 0°, +120°]$ or $[+120°, 0°, -120°]$.

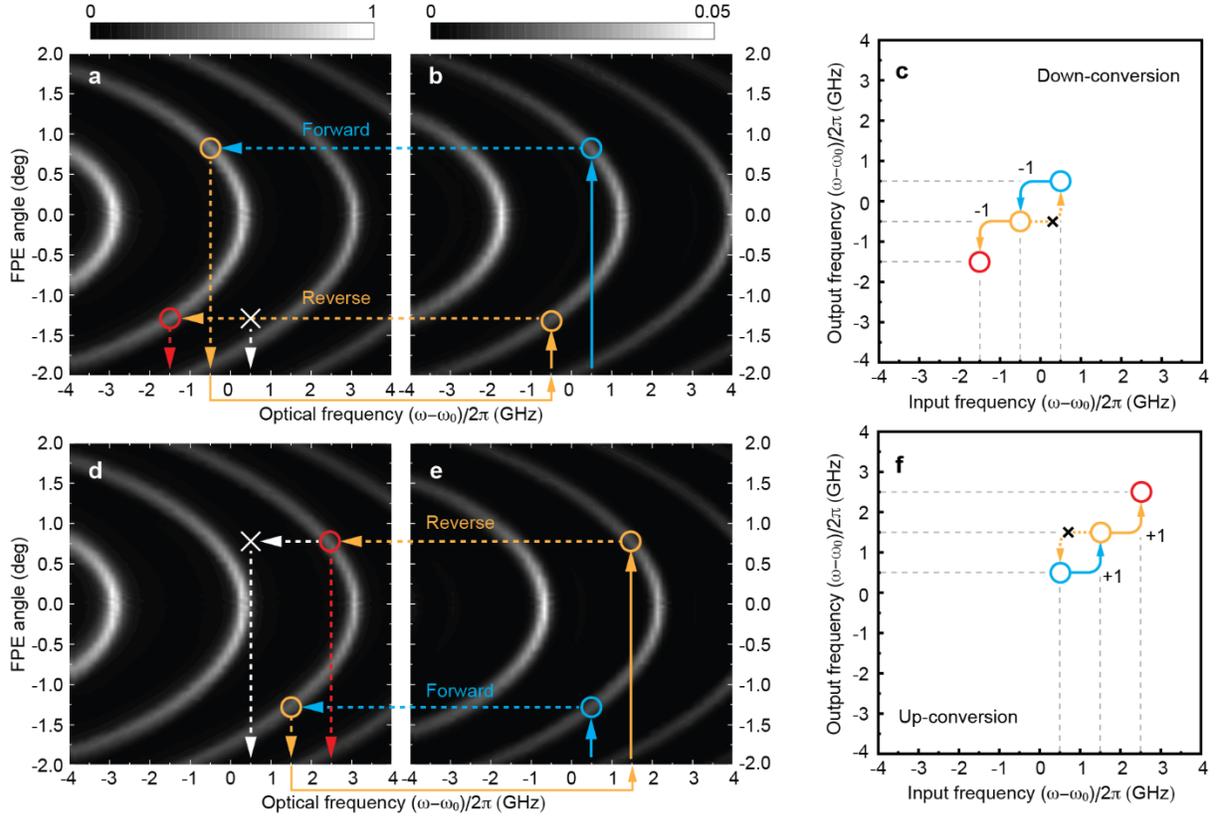

**Fig. 3 | Demonstration of nonreciprocal reflection at mid-IR frequencies from a GHz-modulated STMM.** The 2D maps show the transmission signal through the FPE as a function of the FPE angle and QCL frequency. Panel **(a)** (identical to panel **(d)**) corresponds to the case of the metasurface with no RF spatiotemporal modulation (quasi-DC), and serves as a calibration of the FPE angular position vs frequency expressed as $\omega_0 + \Delta\omega$, where $\omega_0/2\pi$ is some unknown initial optical frequency near 31.6 THz. **(b),** Transmission data for the spatiotemporally modulated metasurface for the modulation sequence $[+120°, 0, -120°]$ corresponding to frequency down-conversion. **(e),** Same for the sequence $[-120°, 0, +120°]$ corresponding to frequency up-conversion. Panels **(c)** and **(f)** depict the two steps of frequency down- and up-conversion. The modulation frequency is $\Omega/2\pi = 1$ GHz.

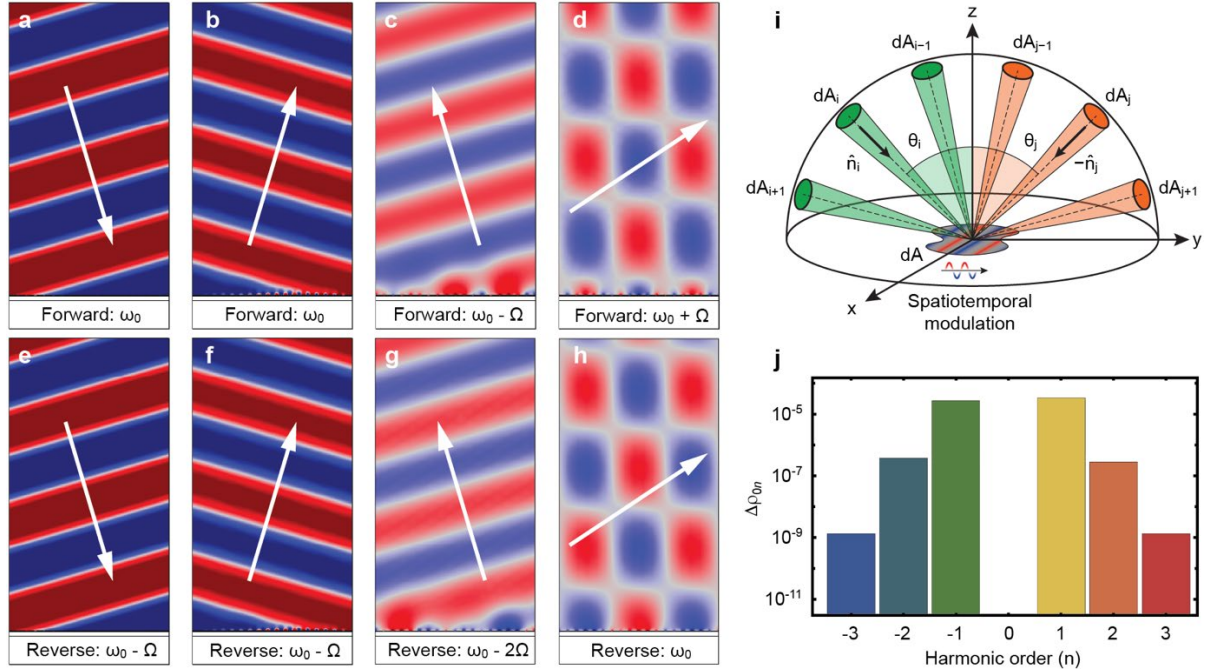

**Fig. 4 (a)** Distribution of the electric field impinging on the spatiotemporally modulated metasurface at the fundamental frequency $\omega_0$ and at an angle $\theta_L = 16.1°$ corresponding to the Littrow condition. The simulated reflected electric field distributions at the **(b)** fundamental, **(c)** down-converted and **(d)** up-converted frequencies. The corresponding plots for the reverse experiment for an incident electromagnetic wave with frequency $\omega_0 - \Omega$ illuminating the metasurface at the Littrow angle is shown for the **(e)** incident field, **(f)** specular reflection, **(g)** down-conversion, and **(h)** up-conversion processes. Owing to the breakdown of Lorentz reciprocity induced by the spatiotemporal modulation, the up-converted field at frequency $\omega_0$ in **(h)** is reflected to a direction $\theta = -56.3° \neq \theta_L$. **(i)** Polka dot pattern of emitters and receivers on a black body enclosure of the STMM for modeling violation of Kirchhoff's law from a spatiotemporally modulated metasurface. An STMM subjected to a travelling-wave modulation is in thermodynamic equilibrium with its unit hemispherical enclosure, which is black over a frequency comb made of narrow bands of angular frequencies $\omega \in [\omega_m, \omega_m + d\omega]$ and corresponding wavevectors $\boldsymbol{k}_m = \boldsymbol{k}_0 + m\boldsymbol{\beta}$ and solid angles, and perfectly reflecting otherwise. **(j)** Contribution of different harmonics to the breakdown of Kirchhoff's law as expressed in Eq. (5) of the main text. Our numerical simulations are in the perturbative regime of small modulation amplitude and frequency: $V_0 = 2$ V, $\Delta V/V_0 = 0.05$, $\omega_0/2\pi = 31.6$ THz, $\Omega/2\pi =$

1 GHz. Other parameters are $\beta = 2\pi/3d = 0.367 \ \mu m^{-1}$, graphene mobility $\mu = 800 \ cm^2/(V \cdot s)$, and capacitance of the metadevice $C = 3.743 \ mF/m^2$.

**Supplementary Information**

**Supplementary Note S1: Sample Fabrication**

Our STMM samples are fabricated on 20 mm by 20 mm, 300 nm-thick thermal oxide-coated intrinsic silicon substrates. First, all electrodes are defined in a resist layer (AZ 5214) by photolithography using a Heidelberg MLA 150 maskless aligner, followed by electron-beam (e-beam) evaporation of Ti/Al/Ti/Au (10 nm/200 nm/15 nm/150 nm) and liftoff. An additional photolithography step using the same photoresist and e-beam evaporation of Ti/Pt (15 nm/70 nm) around the bonding pads and liftoff are performed to facilitate the flip chip bonding. The device active region is then defined by photolithography in a bilayer photoresist stack (LOR 10B/AZ 5214). After a short plasma cleaning step (power = 200 W, time = 1 min), Cr/Au (5 nm/100 nm) and then Cr/a-Ge (5 nm/500 nm) are deposited by DC sputtering (power = 300 W, pressure = 3 mTorr) at room temperature, which serve as the metal ground plane and the cavity layer, respectively. After sputtering, the samples are soaked in resist remover (Remover PG) at 75 °C for 2 hours to strip the bilayer resist (**Fig. S1.1**).

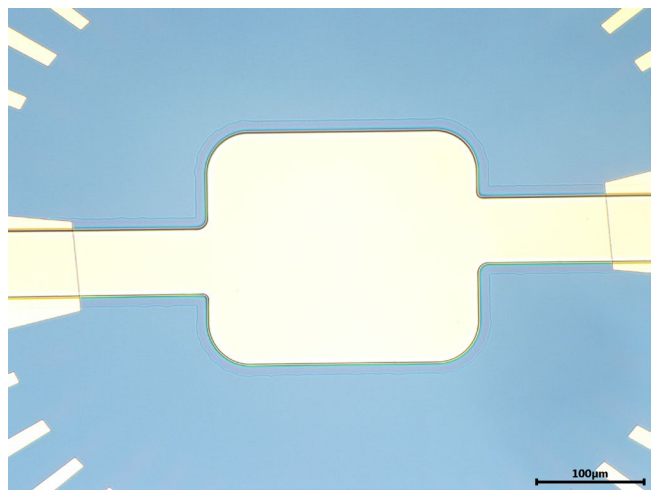

**Fig. S1.1:** Optical image showing STMM device active region after a-Ge mesa formation.

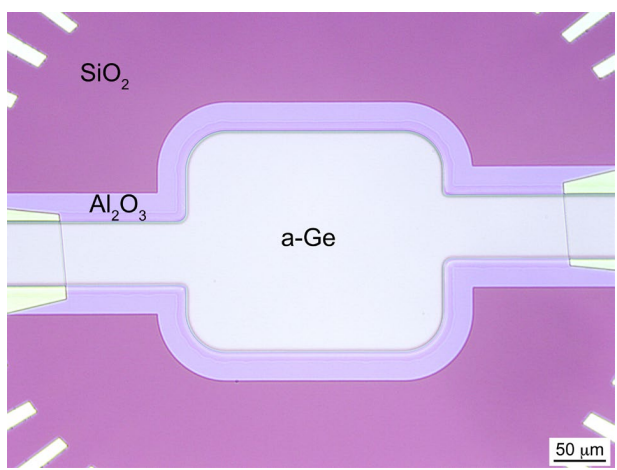

**Fig. S1.2:** Optical image showing STMM device active region after alumina insulation.

After a-Ge mesa formation, a 20 nm thick alumina isolation layer is deposited at 250 °C by atomic layer deposition (ALD) using trimethylaluminum (TMA) and water as precursors. Then, to remove the alumina on top of metal electrodes and around a-Ge mesas, photolithography is performed using AZ 5214 photoresist, which serves as etching mask protecting the a-Ge mesas, followed by an inductively coupled plasma reactive ion etching (ICP-RIE) step with chlorine-based chemistry (ICP power = 350 W, RIE power = 35 W). After alumina etching, the resist etching mask is removed by solvents and oxygen plasma (**Fig. S1.2**).

After alumina isolation is completed, PMMA-coated single-layer graphene purchased from ACS Material is first released on a deionized wafer and then transferred onto the STMM samples. Upon drying in ambient for several hours, the samples are baked on a hotplate at 120 °C for 20 mins to enhance the adhesion of the graphene sheet, and then soaked in room-temperature acetone bath overnight to remove the protective PMMA layer (**Fig. S1.3**).

The transferred graphene sheet is then patterned by photolithography using photoresist (AZ 5214) as etching mask and oxygen plasma etching (100W for 2 mins). The resist mask is cleaned by solvents after etching. Finally, one-dimensional rectangular patch antennas on top of a-Ge mesas and metal connectors that connect the patch antennas to the predefined electrodes on thermal oxide are defined in a bilayer PMMA resist stack (495A4/950A2) by e-beam lithography (JEOL 6300), followed by e-beam evaporation of Au (thickness = 40 nm) and liftoff in room-temperature acetone. The completed STTM device is shown in **Figs. 1** e-g in the main text.

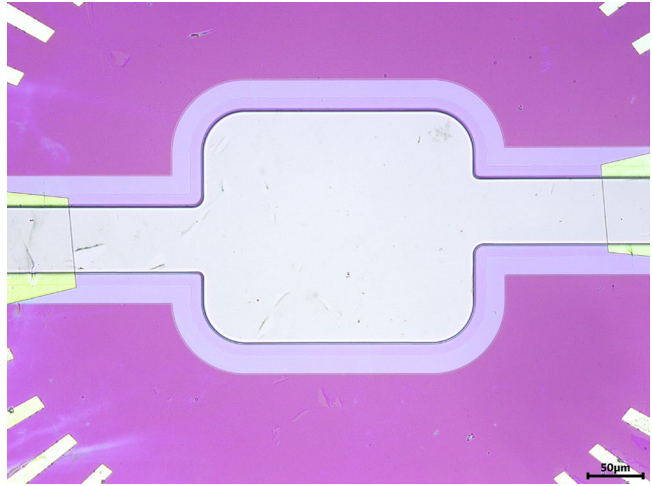

**Fig. S1.3:** Optical image showing STMM device active region after graphene transfer and PMMA removal.

**Supplementary Note S2: Sample Design and COMSOL Simulations**

The base pixel structure of the metasurface was designed to provide optimal performance at both 30 THz (optical) and RF regimes. In the optical domain we require the largest synthetic diffracted signal into order $m = -1$ at the Littrow angle. This roughly corresponds to the maximum optical phase modulation with graphene Fermi energy modulation in a reasonable applied voltage range – modeled from 0 to 0.7 eV. The base pixel is designed to operate in a slightly overcoupled regime [1] where the reflected phasor circles around the origin as a function of both frequency and graphene Fermi energy – see below. At the same time, at RF, we require a minimal RC constant to enable GHz modulation as well as no more than a few volt signal amplitudes to modulate graphene in the largest range possible. The latter requirement, along with pixel size, restricts pixel capacitance to be no smaller than a certain value. Minimization of resistance is also limited by the cross-sectional area of antennas, which also serve as RF contacts.

The theoretically possible corner frequency $f_c = 2\pi/RC$ can reach few hundred MHz, assuming bulk resistivity values for gold as antenna material. It is known, however, that nanostructured metals usually show larger resistivity values [2] leading to lower corner frequencies. Thus, GHz operation of our pixels could only be achieved at the sloping region of the low-pass RC filter curve. Moreover, it was found that interpixel coupling further complicates the pixel's high-frequency RF behavior.

It was found through simulations that a simplified device structure consisting of antennas over a ground plane separated by 10-20 nm of dielectric [3] results in substantially undercoupled performance at 30 THz and so in the final design we inserted a layer of electrically conductive but optically transparent germanium on top of the ground plane. Germanium was chosen instead of silicon for the reason of easier identification and differentiation from the substrate in energy-dispersive X-ray (EDX) characterization of the sample. A 20 nm alumina layer was placed on top of germanium, followed by graphene. Thus, the graphene-alumina-germanium comprise the capacitor to be charged-discharged at GHz frequency for Fermi energy modulation in graphene. This structure, along with the ground plane under germanium also operates as a patch antenna at 10 $\mu$m optical frequency with characteristic resonant distribution of electric field concentrated mostly in a lower-index dielectric (alumina) as is typical for hybrid metal-dielectric-semiconductor waveguides [4].

Once the basic structure of the device was identified, a 2D COMSOL model was created for one synthetic pixel consisting of three base pixels. Each base pixel contained six gold strips operating as 1D patch antennas. A two-parameter sweep was performed with antenna width and germanium thickness as parameters in a $m = -1$ diffraction order in Littrow configuration (see left panel in **Fig. S2.1**). Peak fractional power values in the synthetic $m = -1$ order approaching 1% were obtained in the simulations. Importantly, the range of parameters yielding relatively high diffracted powers was found to be substantially wide allowing for moderate fabrication errors. Following the optimization, we selected 840 nm for antenna width and 460 nm for Ge thickness and performed a frequency sweep for a single base pixel computing the optical reflected phasor in order to identify the regime of operation of the device (see left panel in **Fig. S2.2**). Unlike the specular reflection, the synthetic diffraction signal shows a near-linear phase evolution of $2\pi$ range over one period of modulation frequency. This translates into a frequency upshift by one Fourier harmonic, as is seen on the right panel in **Fig. S2.2**. The smaller peaks observed on the Fourier spectrum are due to remaining residual nonlinearity in the phase evolution of $S_{21}$ as well as oscillations on the amplitude function.

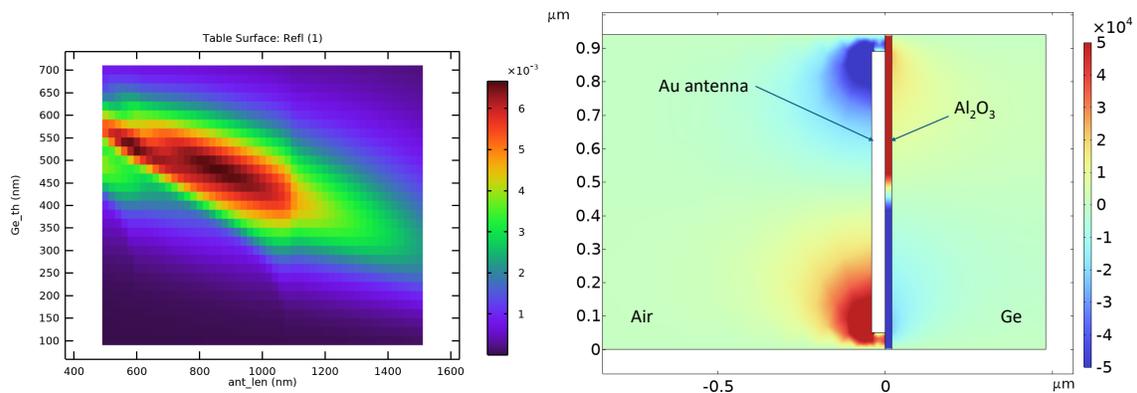

**Fig. S2.1.** Left: Fractional power of $m = -1$ synthetic diffraction at Littrow angle for the 3-pixel periodic device as a function of antenna size and germanium thickness. Alumina thickness is 20 nm. Optical frequency is 30 THz. Graphene Fermi energies are set to 0.2 eV, 0.2 eV, and 0.5 eV to the three base pixels, and scattering rate is 10 fs. Right: distribution of the x-component (i.e., along the vertical direction) of electric field around the antenna (units: V/m).

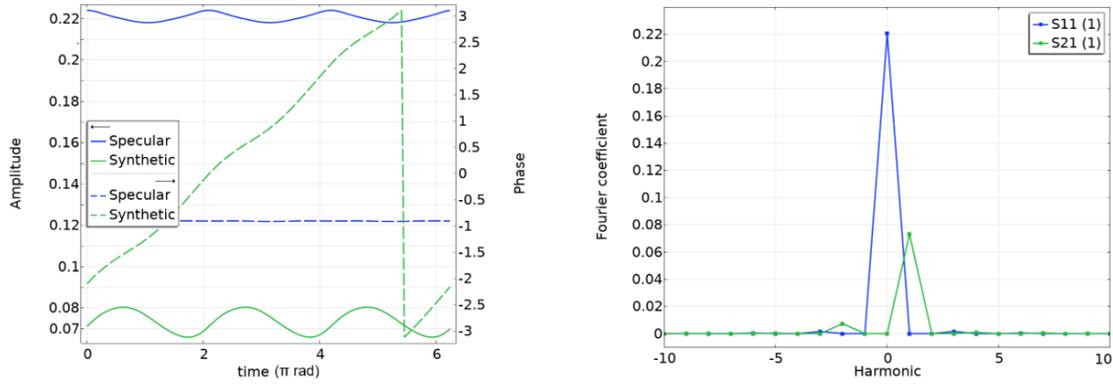

**Fig. S2.2.** Left: Time evolution of specular (blue) and synthetic (green) amplitude (solid line) and phase (dashed line) over a single oscillation period of 1 GHz modulation frequency implemented as graphene Fermi energy modulation applied to base pixels $n = 1, 2,$ and 3 as $E_{F,n} = E_{F,DC} + E_{F,ampl} \cos\left(t_n + [n-1]\frac{2\pi}{3}\right)$, where $E_{F,DC} = 0.3$ eV, and $E_{F,ampl} = 0.2$ eV. Right: Amplitude coefficients of the Discrete Fourier transform of the complex signals from the left panel shown as a function of Fourier harmonic number for the specular (blue) and synthetic (green) signals $S_{11}$ and $S_{21}$ respectively.

We also evaluated the response of a single base pixel, or rather a periodic array of identical base pixels, as a function of frequency and graphene Fermi energy. **Fig. S2.3** shows specular reflection power and phase of $S_{11}$ parameter at zero incidence angle as plots of power and phase, as well as a phasor, as a function of graphene Fermi energy at 30 THz optical frequency. A phase modulation range in excess of 180° can be seen. The polar plot clearly shows the overcoupled regime. Results obtained at non-zero angles of incidence are very similar to normal incidence. **Fig. S2.3** similarly shows dependence on optical frequency, where optical phonon in alumina is clearly seen at 21.6 THz.

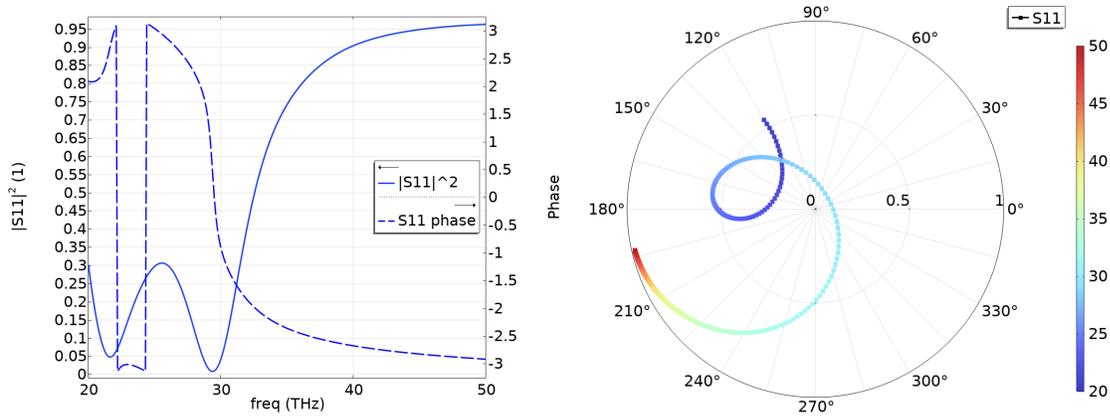

**Fig. S2.3.** Specular reflection from a periodic array of base pixels at $E_F = 0.2$ eV as a function of optical frequency shown as power and phase (Left panel) and as a phasor polar plot (Right panel). The dip at 21.5 THz is due to the optical phonon in alumina.

The polarization response of the device, experimentally measured by an FTIR microscope (Hyperion 2000), is compared to the COMSOL simulation results in **Fig. S2.4**. The width of the experimentally measured resonance is broader than the simulated one due to the finite angular acceptance cone of the FTIR microscope. The resonance frequency of the device is somewhat closer to the 30 THz frequency, so we perform our main experiments at a slightly higher frequency around 9.8 $\mu$m in the overcoupled regime where the synthetic diffraction signal is maximized.

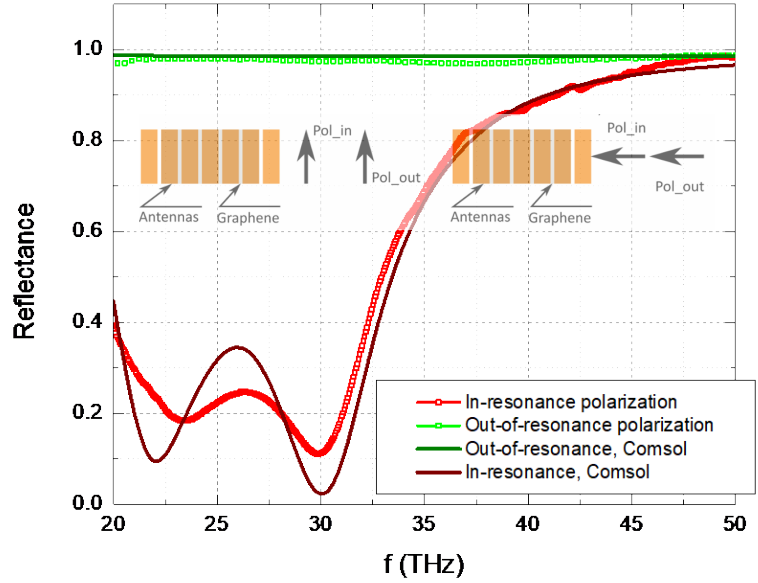

**Fig. S2.4.** Experimentally measured (symbols) and COMSOL results (solid lines) for in-resonance (red) and out-of-resonance (green) reflectivity from the device. Inset shows input and output polarization orientations with respect to antennas for in-resonance (right side) and out-of-resonance (left side) orientations.

**Supplementary Note S3: Driver Electronics**

A single RF synthesizer (Hewlett Packard 6082A) was used to generate a sinusoidal signal at 1 GHz and adjustable amplitude. The signal is fed into a custom-build 3-channel driver PCB featuring three independent voltage-controlled phase shifters (Mini Circuits SPHSA-152+). The control voltages for the phase shifters are supplied from two AO channels of a DAQ unit (NI USB-6363) and are set in a custom-written LabView program. Only two voltages are needed to control the two-phase differences between three phase shifters. The voltage-phase calibrations for the phase shifters were separately obtained using a vector network analyzer (Anritsu MS2036C) at 1 GHz operating frequency. The three outputs from the phase shifters are fed into three RF power amplifiers (Mini Circuits ZHL-10W-2G+) with three RF isolators (DiTom Microwave D3I0810S) connected at their outputs to absorb back-reflected RF signals. The three amplified and phase-shifted RF signals are fed into the sum ports of three $1 \times 12$ RF power splitters (Mini Circuits ZN12PD-252-S+). The power splitters feature the standard SMA connectors, total number being 36. We employ 36 individual coax cables of identical length to run the signals to an adapter PCB, the purpose of which is to transfer the RF signals from bulky cables to a pair of micro-coax ribbons (Samtec EQCD-020-24.00-SBR-TTR-1), which are then run to the sample PCB.

The metasurface sample is flip-chip mounted on a small 40 mm $\times$ 40 mm sample PCB, which has a through hole for optical access to the active area of the device and two receptacle sockets for the micro-coax ribbons located on the opposite sides of the sample. The sample is designed so that contacts for the odd- and even-numbered pixels are located on the opposite edges of the chip and are served by the respective micro-coax ribbons. All connections and PCB microstrip waveguides up to the sample contacts are designed for 50 Ohm impedance; however, the sample is obviously not impedance matched for reflectionless power transfer, which necessitates the use of RF isolators in the signal chain. Despite good design efforts, some interpixel crosstalk is present due in part to the proximity of the pixels on the device, but mostly due to the micro-coax connectors. This crosstalk is believed to be substantially responsible for relatively low sample diffraction efficiency.

**Supplementary Note S4: Simulations of Nonreciprocal Scattering in STMMs**

To computationally demonstrate the breakdown of Lorentz reciprocity in light scattering with our spatiotemporally modulated metasurface, we perform full wave simulations in COMSOL Multiphysics. Optical properties of the Ge, Al$_2$O$_3$, and Au domains were extracted from the literature [5–7]. The reflected field amplitudes are obtained by numerically solving Maxwell's equations in frequency domain for each Floquet harmonic while modeling the graphene monolayer as an induced surface current density,

$$\boldsymbol{J}(\boldsymbol{R},\omega) = \int_{-\infty}^{\infty} \boldsymbol{J}(\boldsymbol{R},t)e^{i\omega t}dt = \int_{-\infty}^{\infty} e^{i\omega t}dt \int d\boldsymbol{R}' \int \sigma(\boldsymbol{R},\boldsymbol{R}';t,t')\boldsymbol{E}(\boldsymbol{R}',t')dt', \quad \text{(S.1)}$$

where $\boldsymbol{E}(\boldsymbol{R},t)$ is the electric field amplitude on the monolayer and $\sigma(\boldsymbol{R},\boldsymbol{R}';t,t')$ is graphene's linear response function. In the absence of spatiotemporal modulation and by neglecting spatial dispersion,

$$\sigma(\boldsymbol{R},\boldsymbol{R}';t,t') = \delta(\boldsymbol{R}-\boldsymbol{R}')\sigma_{un}(t-t')$$

$$= \delta(\boldsymbol{R}-\boldsymbol{R}')\frac{1}{2\pi}\int_{-\infty}^{\infty}\sigma_{un}(\omega')e^{-i\omega'(t-t')}d\omega', \quad \text{(S.2)}$$

resulting in an unmodulated surface current density given by

$$\boldsymbol{J}_{un}(\boldsymbol{R},\omega) = \sigma_{un}(\omega)\boldsymbol{E}(\boldsymbol{R},\omega). \quad \text{(S.3)}$$

Here,

$$\sigma_{un}(\omega) = \sigma_0 \left\{ \frac{i}{\pi} \frac{4E_F^{un}}{\hbar(\omega_i)} + \Theta(\hbar\omega_i - 2E_F^{un}) + \frac{i}{\pi}\log\left|\frac{\hbar\omega_i - 2E_F^{un}}{\hbar\omega_i + 2E_F^{un}}\right| \right\} \quad \text{(S.4)}$$

is the complex zero-temperature optical conductivity of graphene accounting for both intraband and interband transitions [8] with $\omega_i = \omega + i\gamma$. Also, $\sigma_0 = e^2/4\hbar$ is graphene's universal conductivity, $\gamma^{-1} = \mu E_F^{un}/(e^2 v_F^2)$ is a phenomenological relaxation time, $\mu$ is the mobility of charge carries (assumed to be 800 cm²/V · s in our simulations), and $v_F = 10^6$ m/s is the Fermi velocity. Finally, $E_F^{un}$ is the doping level of graphene which depends on the applied gate voltage $V_0$ (= 2 V in the simulations) as [9]

$$E_F = \hbar v_F \sqrt{\frac{\pi C V}{e}}, \tag{S.5}$$

where $C = 3.743$ mF/m² is the capacitance per unit of area, which we compute via $C = 2U/(A V^2)$, where $U$ the energy density stored in the metasurface of area $A$ when subjected to an electrostatic potential $V$.

To model the influence of a spatiotemporally varying gate voltage

$$V(\mathbf{R}, t) = V_0 + \Delta V \cos(\Omega t - \boldsymbol{\beta} \cdot \mathbf{R}) \tag{S.6}$$

on the system's optical response, we assume that the charge carriers respond adiabatically and locally to this modulation. In this case, graphene's linear response function can be cast as

$$\sigma(\mathbf{R}, \mathbf{R}'; t, t') = \delta(\mathbf{R} - \mathbf{R}')[\sigma_{un}(t - t') + \Delta\sigma(t - t') \cos(\Omega t - \boldsymbol{\beta} \cdot \mathbf{R})], \tag{S.7}$$

implying an effective conductivity at frequency ω,

$$\sigma(\mathbf{R}, \omega; t) = \sigma_{un}(\omega) + \Delta\sigma(\omega) \cos(\Omega t - \boldsymbol{\beta} \cdot \mathbf{R}). \tag{S.8}$$

The resulting modulated surface current density at frequency ω is then given by

$$\mathbf{J}_m(\mathbf{R}, \omega) = \sigma_{un}(\omega)\mathbf{E}(\mathbf{R}, \omega)$$
$$+ \frac{1}{2}\left[\Delta\sigma(\omega - \Omega)e^{i\boldsymbol{\beta}\cdot\mathbf{R}}\mathbf{E}(\mathbf{R}, \omega - \Omega) + \Delta\sigma(\omega + \Omega)e^{-i\boldsymbol{\beta}\cdot\mathbf{R}}\mathbf{E}(\mathbf{R}, \omega + \Omega)\right]. \tag{S.9}$$

This expression governs the coupling between harmonics and is supplied as a boundary condition current density for each individual harmonic in our COMSOL simulations. Here, $\Delta\sigma(\omega)$ is a frequency dispersive complex function that determines the amplitude of the reflected up and down converted harmonics in the system.

An expression for $\Delta\sigma(\omega)$ can be derived by noticing when the voltage modulation amplitude is much smaller than the onset potential ($\Delta V \ll V_0$), and in the adiabatic regime assumed here, the instantaneous and spatially local doping level is obtained by substituting Eq. (S.6) into Eq. (S.5):

$$E_F(\mathbf{R}, t) \approx E_F^{un}\left[1 + \frac{\Delta V}{2V_0}\cos(\Omega t - \boldsymbol{\beta}\cdot\mathbf{R})\right]. \tag{S.10}$$

Finally, we obtain the form of the conductivity in Eq. (S.8) by substituting the expression from Eq. (S.10) into Eq. (S.4) and expanding it to linear order in $\Delta V$. This procedure allows us to identify

$$\Delta\Delta\sigma(\omega) = \frac{2i\,\sigma_0 E_F^{un}}{\pi\hbar(\omega + i\gamma)}\frac{\Delta V/V_0}{1 + \hbar^2(\omega + i\gamma)^2/E_F^{un\,2}} \tag{S.11}$$

In our simulations we consider that the modulation amplitude is $\Delta V/V_0 = 0.05$, consistent with our perturbative approach. Although in our experimental demonstration we apply a much higher modulation voltage ($\Delta V/V_0 \approx 1$), the large RC time constant of our device significantly reduces effective modulation voltage at the device, as evidenced by the qualitative agreement between the simulated and measured mode conversion efficiencies.

**Supplementary Note S5: Theory of Nonreciprocal Scattering and Breakdown of Kirchhoff's Law**

Here, we show in detailed steps that nonreciprocal frequency conversion upon reflection from a spatiotemporally modulated metasurface (STMM) is equivalent to the violation of Kirchhoff's law of thermal radiation by extending prior work in the context of magneto-optic materials [10–13]. We also look at a number of limiting cases to confirm that our key result reduces to previously reported versions of Kirchhoff's law of thermal radiation.

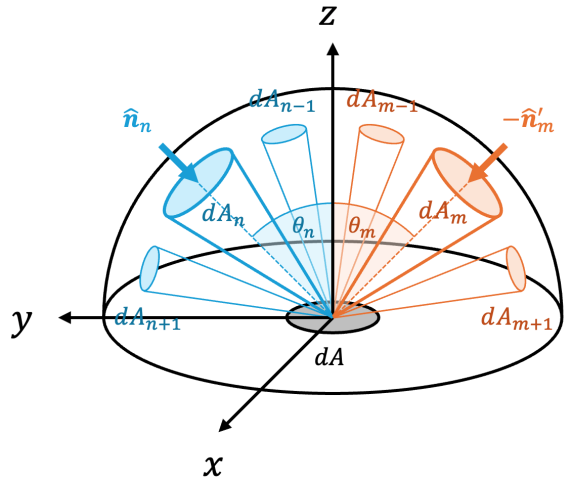

**Fig. S5.1. Problem definition.** An opaque graybody $dA$ is in thermodynamic equilibrium with its unit hemispherical enclosure, which is black over a frequency comb made of narrow bands of angular frequencies $\omega \in [\omega_m, \omega_m + d\omega]$ and perfectly reflecting otherwise. Each $\omega_m = \omega_0 + m\Omega$ and has a corresponding wavevector $\boldsymbol{k}_m = \boldsymbol{k}_0 + m\boldsymbol{\beta}$ and differential area $dA_m$ and solid angle, resulting in a "polka dot pattern" of emitting and absorbing areas on the enclosure. The graybody emits, absorbs, and reflects light along direction vectors $\hat{\boldsymbol{n}} = \sin\theta\cos\phi\,\hat{\boldsymbol{x}} + \sin\theta\sin\phi\,\hat{\boldsymbol{y}} + \cos\theta\,\hat{\boldsymbol{z}} = c\boldsymbol{k}/\omega$, where $\theta$ and $\phi$ are polar and azimuthal angles of incidence and $\boldsymbol{k}$ is the wavevector of light. The "reflected direction vector" is defined as $\hat{\boldsymbol{n}}' = \sin\theta\cos\phi\,\hat{\boldsymbol{x}} + \sin\theta\sin\phi\,\hat{\boldsymbol{y}} - \cos\theta\,\hat{\boldsymbol{z}} = c\boldsymbol{k}'/\omega$.

First, let us reiterate the problem definition in the main text. Consider an opaque, spatiotemporally modulated graybody $dA$ in thermodynamic equilibrium with its unit hemispherical enclosure, which is black over a frequency comb made of narrow bands of angular frequencies $\omega \in [\omega_m, \omega_m + d\omega]$ and perfectly reflecting otherwise. Each $\omega_m = \omega_0 + m\Omega$ ($m \in \mathbb{Z}$) and has a corresponding longitudinal wavevector (i.e., in the $xy$-plane) $\boldsymbol{k}_{\parallel,m} = \boldsymbol{k}_{\parallel,0} + m\boldsymbol{\beta}$ as well as differential area $dA_m$ and solid angles $d\Omega_{dA \to dA_m}$ and $d\Omega_{dA_m \to dA}$. This results in a "polka dot pattern" of emitting and receiving areas on the enclosure (see **Fig. 4i** in the main text and **Fig. S5.1** in the Supplementary Information). In other words, the enclosure perfectly emits and absorbs the electromagnetic modes scattered by the graybody that are a result of spatiotemporal modulation, where $\Omega$ is the modulation frequency and $\boldsymbol{\beta}$ is the gradient of the phase profile (e.g., on the surface of the graybody via a spatiotemporally modulated metasurface) [14–16] The graybody emits, absorbs, and reflects light along direction vectors $\hat{\boldsymbol{n}}_m = \sin\theta_m \cos\phi_m\, \hat{\boldsymbol{x}} + \sin\theta_m \sin\phi_m\, \hat{\boldsymbol{y}} + \cos\theta_m\, \hat{\boldsymbol{z}} = c\boldsymbol{k}_m/\omega_m$, where $\theta_m$ and $\phi_m$ are polar and azimuthal angles of incidence associated with spectral-directional channel $m$, and $\boldsymbol{k}_m = \boldsymbol{k}_{\parallel,m} + k_{z,m}\hat{\boldsymbol{z}}$ is the free-space wavevector of light (also, $k_{zm} = \sqrt{(\omega_m/c)^2 - |\boldsymbol{k}_{\parallel,m}|^2}$). We define the reflected direction vector as well, in which the sign of the $z$-component is flipped: $\hat{\boldsymbol{n}}'_m = \sin\theta_m \cos\phi_m\, \hat{\boldsymbol{x}} + \sin\theta_m \sin\phi_m\, \hat{\boldsymbol{y}} - \cos\theta_m\, \hat{\boldsymbol{z}} = c\boldsymbol{k}'_m/\omega_m$. The relationships between the direction vectors $\hat{\boldsymbol{n}}_m$ and $\hat{\boldsymbol{n}}'_m$ as well as their flipped-sign counterparts are schematized in **Fig. S5.2**, along with solid angles associated with them (to be discussed). The schematic from the main text illustrating our "polka dot pattern" approach is reproduced in **Fig. S5.1**. Again, our goal is to obtain a relationship between the light emitted and absorbed by the graybody, even when spectral directional emissivity does not equal spectral directional absorptivity (i.e., the original form Kirchhoff's law of thermal radiation breaks down). We will work in the perturbative regime of small modulation frequency and amplitude, as well as assume the system does not exchange energy with the source of modulation. We will work in the perturbative regime of small modulation frequency and amplitude and assume the system does not exchange energy with the modulation source.

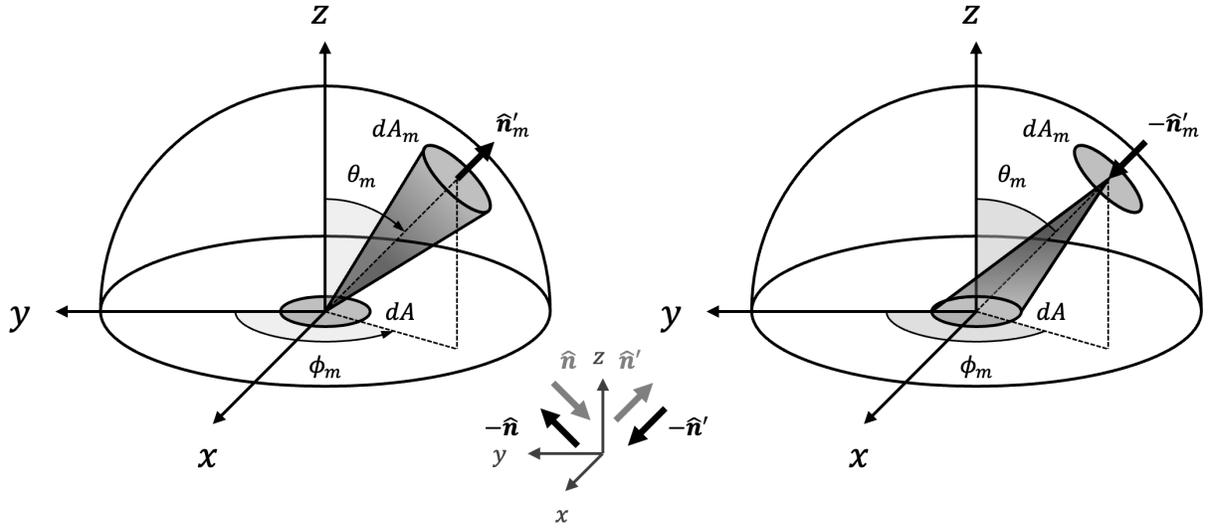

**Fig. S5.2: Geometric definitions.** (Left) Definition of the solid angle $d\Omega_{dA \to dA_m}$. For a unit hemisphere, $d\Omega_{dA \to dA_m} = dA_m \cos\theta_m$. (Right) Definition of the solid angle $d\Omega_{dA_m \to dA} = dA \cos\theta_m$. As a result of reciprocity of the geometric view factor, $dA\, d\Omega_{dA \to dA_m} = dA_m\, d\Omega_{dA_m \to dA}$. (Bottom center) The relationships between the direction vectors $\hat{n} = \sin\theta \cos\phi\, \hat{x} + \sin\theta \sin\phi\, \hat{y} + \cos\theta\, \hat{z}$ and $\hat{n}' = \sin\theta \cos\phi\, \hat{x} + \sin\theta \sin\phi\, \hat{y} - \cos\theta\, \hat{z}$, as well as their flipped-sign counterparts.

**Emission equation:** First, consider the case of light that leaves $dA$ and arrives at all possible "receiver polka dots" or differential areas $dA_m$ on the enclosure. This includes (1) light emitted from $dA$, (2) light from all possible "emitter polka dots" or $dA_n$'s reflected off $dA$ in the direction of the $dA_m$'s, and (3) light emitted by the rest of the enclosure. Because of the geometric reciprocity of the diffuse view factor ($A_i F_{ij} = A_j F_{ji}$, where $F_{ij}$ is the diffuse view factor from $A_i$ to $A_j$, which is not to be confused with Lorentz reciprocity itself), the third contribution is zero [13,17]. For the first contribution, the radiant power of emitted light accounting for all possible modes $m$ is given by

$$\sum_m e(\omega_m, \hat{\boldsymbol{n}}'_m) I_b(\omega_m, T) dA d\Omega_{dA \to dA_m}, \tag{S.12}$$

where $e(\omega_m, \hat{\boldsymbol{n}}'_m)$ is the spectral directional emissivity, $I_b(\omega_m, T)$ is the blackbody spectral radiance, and $d\Omega_{dA \to dA_m}$ is the solid angle subtended by $dA_m$ viewed from $dA$. In general, solid angle is defined by $d\Omega = dA \cos\theta / r^2$ (where $dA$ is an general differential area being viewed, not the graybody), but since the enclosure is a unit hemisphere, $d\Omega_{dA \to dA_m} = dA_m \cos\theta_m$. Therefore, the radiant power emitted by $dA$ is

$$\sum_m e(\omega_m, \hat{\boldsymbol{n}}'_m) I_b(\omega_m, T) dA dA_m \cos\theta_m. \tag{S.13}$$

For the second contribution, the radiant power entering $dA$ is given by

$$\sum_n I_b(\omega_n, T) dA_n d\Omega_{dA_n \to dA} \tag{S.14}$$

accounting for all possible modes $n$ once again. From the definition of solid angle, $d\Omega_{dA_n \to dA} = dA \cos\theta_n$. Therefore, Eq. (S.14) becomes

$$\sum_n I_b(\omega_n, T) dA dA_n \cos\theta_n. \tag{S.15}$$

This radiant power, emitted by $dA_n$, is reflected by $dA$. If the graybody is spatiotemporally modulated, the reflected light can scatter into multiple spectral-directional channels. As previously mentioned, the reflected light will consist of multiple modes with frequency and spatial harmonics $\omega_m = \omega_0 + m\Omega$ and $\boldsymbol{k}_m = \boldsymbol{k}_0 + m\boldsymbol{\beta}$, where $\Omega$ is the modulation frequency and $\boldsymbol{\beta}$ is the phase gradient. Therefore, differently from prior work [11,13], we define the bidirectional reflectance distribution function (BRDF) on the basis of both incident and reflected frequency and direction vector: $\rho(\omega_{inc} \to \omega_{ref}, \hat{\boldsymbol{n}}_{inc} \to \hat{\boldsymbol{n}}_{ref})$. Therefore, for light emitted by all possible $dA_n$'s that is scattered into all possible spectral-directional channels $m$, the radiant power of the reflected light can be written as

$$\sum_m \left[ \sum_n \rho(\omega_n \to \omega_m, \hat{\boldsymbol{n}}_n \to \hat{\boldsymbol{n}}'_m) I_b(\omega_n, T) dA dA_n \cos\theta_n \right] dA_m \cos\theta_m. \quad (S.16)$$

Here, $\hat{\boldsymbol{n}}_n$ is the direction vector associated with light emitted by a particular $dA_n$ at angular frequency $\omega_n$, and we have used the fact that $d\Omega_{dA \to dA_m} = dA_m \cos\theta_m$. Since the BRDF is summed over all possible $dA_n$'s (integrated over the hemisphere in the limit as $d\Omega_{dA_n \to dA} \to 0$), it accounts for retroreflection, i.e., from $dA_m$ back to $dA_m$ itself. The sum of Eqs. (S.13) and (S.16) constitute the radiant power arriving at all possible $dA_m$'s. This must be balanced by the radiant power leaving all possible $dA_m$'s and heading toward $dA$:

$$\sum_m I_b(\omega_m, T) dA dA_m \cos\theta_m \quad (S.17)$$

where we have once again used the fact that $d\Omega_{dA_m \to dA} = dA \cos\theta_m$. Therefore,

$$\sum_m I_b(\omega_m, T) dA dA_m \cos\theta_m = \sum_m e(\omega_m, \hat{\boldsymbol{n}}'_m) I_b(\omega_m, T) dA dA_m \cos\theta_m$$

$$+ \sum_m \left[ \sum_n \rho(\omega_n \to \omega_m, \hat{\boldsymbol{n}}_n \to \hat{\boldsymbol{n}}'_m) I_b(\omega_n, T) dA dA_n \cos\theta_n \right] dA_m \cos\theta_m$$

$$\Rightarrow 0 = \sum_m I_b(\omega_m, T) dA dA_m \cos\theta_m \left\{ 1 - e(\omega_m, \widehat{\boldsymbol{n}}'_m) \right.$$

$$\left. - \sum_n \rho(\omega_n \to \omega_m, \widehat{\boldsymbol{n}}_n \to \widehat{\boldsymbol{n}}'_m) \frac{I_b(\omega_n, T)}{I_b(\omega_m, T)} dA_n \cos\theta_n \right\}. \quad \text{(S.18)}$$

The term inside the curly brackets is a nondimensionalized representation of the radiant power leaving all possible $dA_m$'s minus the emitted and reflected radiant powers arriving from $dA$. This term must be greater than or equal to zero. If it were less than zero, it would imply that the "receiver polka dots" are emitting more energy than they are absorbing, violating the second law of thermodynamics. However, if the summation over $m$ is equal to zero and the term in the curly brackets must be nonnegative, then

$$0 = 1 - e(\omega_m, \widehat{\boldsymbol{n}}'_m) - \sum_n \rho(\omega_n \to \omega_m, \widehat{\boldsymbol{n}}_n \to \widehat{\boldsymbol{n}}'_m) \frac{I_b(\omega_n, T)}{I_b(\omega_m, T)} dA_n \cos\theta_n. \quad \text{(S.19)}$$

**Absorption equation:** Now consider the case of light that arrives at $dA$ after leaving all possible $dA_m$'s. Like the previous case, this includes (1) light absorbed by $dA$, (2) light from all possible $dA_m$'s reflected off $dA$, and (3) light absorbed by the rest of the enclosure. Once again, because of the geometric reciprocity of the diffuse view factor, the third contribution is zero. The rate of absorption is the sum of the radiant powers emitted by each $dA_m$, $I_b(\omega_m, T)dAdA_m \cos\theta_m$, times the spectral directional absorptivity of the graybody:

$$\sum_m a(\omega_m, -\widehat{\boldsymbol{n}}'_m) I_b(\omega_m, T) dA dA_m \cos\theta_m. \quad \text{(S.20)}$$

Similarly to Eq. (S.16), for light emitted by $dA_m$ that is scattered into a particular spectral-directional channel $n$ subtended by the solid angle $d\Omega_{dA \to dA_n} = dA_n \cos\theta_n$, the radiant power of the reflected light can be written as

$$\sum_n \left[ \sum_m \rho(\omega_m \to \omega_n, -\widehat{\boldsymbol{n}}'_m \to -\widehat{\boldsymbol{n}}_n) I_b(\omega_m, T) dA dA_m \cos\theta_m \right] dA_n \cos\theta_n, \qquad (S.21)$$

Therefore, the sum of Eqs. (S.20) and (S.21) is the radiant power arriving at $dA$, and it must be balanced by the radiant power leaving $dA_m$:

$$\sum_m I_b(\omega_m, T) dA dA_m \cos\theta_m = \sum_m a(\omega_m, -\widehat{\boldsymbol{n}}'_m) I_b(\omega_m, T) dA dA_m \cos\theta_m$$

$$+ \sum_m \left[ \sum_n \rho(\omega_m \to \omega_n, -\widehat{\boldsymbol{n}}'_m \to -\widehat{\boldsymbol{n}}_n) I_b(\omega_m, T) dA dA_n \cos\theta_n \right] dA_m \cos\theta_m$$

$$\Rightarrow 0 = \sum_m I_b(\omega_m, T) dA dA_m \cos\theta_m \left\{ 1 - a(\omega_m, -\widehat{\boldsymbol{n}}'_m) \right.$$

$$\left. - \sum_n \rho(\omega_m \to \omega_n, -\widehat{\boldsymbol{n}}'_m \to -\widehat{\boldsymbol{n}}_n) dA_n \cos\theta_n \right\}, \qquad (S.22)$$

where obviously we changed the order of summation in Eq. (S.21) in the first line. Once again, the term inside the curly brackets must be greater than or equal to zero. Otherwise, it would imply $dA$ is absorbing and reflecting more energy than all possible $dA_m$'s are emitting. Therefore, if the summation over $m$ in Eq. (S.22) is equal to zero, then all terms in the summation must be equal to zero as well:

$$0 = 1 - a(\omega_m, -\widehat{\boldsymbol{n}}'_m) - \sum_n \rho(\omega_m \to \omega_n, -\widehat{\boldsymbol{n}}'_m \to -\widehat{\boldsymbol{n}}_n) dA_n \cos\theta_n. \qquad (S.23)$$

**Generalized Kirchhoff's law of thermal radiation for STMMs:** By subtracting Eq. (S.19) from Eq. (S.23), we obtain a relationship between the spectral directional emissivity $e(\omega_m, \widehat{\boldsymbol{n}}'_m)$ and spectral directional absorptivity $a(\omega_m, -\widehat{\boldsymbol{n}}'_m)$:

$$e(\omega_m, \widehat{\boldsymbol{n}}'_m) - a(\omega_m, -\widehat{\boldsymbol{n}}'_m)$$

$$= \sum_n \left[ \rho(\omega_m \to \omega_n, -\widehat{\boldsymbol{n}}'_m \to -\widehat{\boldsymbol{n}}_n) - \rho(\omega_n \to \omega_m, \widehat{\boldsymbol{n}}_n \to \widehat{\boldsymbol{n}}'_m) \frac{I_b(\omega_n, T)}{I_b(\omega_m, T)} \right] dA_n \cos\theta_n \quad (S.24)$$

Equation (S.24) states the following: $e(\omega_m, \widehat{\boldsymbol{n}}'_m)$ does not equal $a(\omega_m, -\widehat{\boldsymbol{n}}'_m)$—meaning Kirchhoff's law of thermal radiation is violated—and the difference between them depends on the difference between the BRDFs along opposite trajectories (i.e., $\widehat{\boldsymbol{n}}_n \to \widehat{\boldsymbol{n}}'_m$ and $-\widehat{\boldsymbol{n}}'_m \to -\widehat{\boldsymbol{n}}_n$) summed over all possible frequency conversions from $\omega_m$ to $\omega_n$ and vice versa. The dependence of $e(\omega_m, \widehat{\boldsymbol{n}}'_m) - a(\omega_m, -\widehat{\boldsymbol{n}}'_m)$ on the directional asymmetry of forward and backward scattering is not a new result [10–13], but the summation over all possible modes $n$ that mode $m$ can scatter into is. Furthermore, the BRDF associated with backward scattering, $\rho(\omega_n \to \omega_m, \widehat{\boldsymbol{n}}_n \to \widehat{\boldsymbol{n}}'_m)$, is scaled by $I_b(\omega_n, T)/I(\omega_m, T)$. This bears some similarities to generalized reciprocity in time-modulated systems, which states that the Green's function of forward scattering is related to the Green's function of backward scattering multiplied by a ratio of frequencies [18,19], or the Manly-Rowe relation in nonlinear optics [20].

**Limiting cases:** As a way to check Eq. (S.24), we impose constraints on it until we recover the "original version" of Kirchhoff's law of thermal radiation [17,21]. There are three possible constraints:

(1) no temporal modulation ($\bar{\bar{\varepsilon}} \neq \bar{\bar{\varepsilon}}(t), \bar{\bar{\mu}} \neq \bar{\bar{\mu}}(t)$);

(2) no spatial modulation ($\bar{\bar{\varepsilon}} \neq \bar{\bar{\varepsilon}}(\boldsymbol{r}), \bar{\bar{\mu}} \neq \bar{\bar{\mu}}(\boldsymbol{r})$); and

(3) symmetric permittivity and permeability tensors ($\bar{\bar{\varepsilon}} = \bar{\bar{\varepsilon}}^T, \bar{\bar{\mu}} = \bar{\bar{\mu}}^T$).

All permutations of these constraints are summarized in Table 1 at the end of this section.

*No temporal modulation:* If the graybody is not temporally modulated, $n = m$ because there is no frequency conversion. Therefore, equation (S.24) becomes

$$e(\omega_m, \widehat{\boldsymbol{n}}'_m) - a(\omega_m, -\widehat{\boldsymbol{n}}'_m) = \sum_m \rho(\omega_m \to \omega_m, -\widehat{\boldsymbol{n}}'_m \to -\widehat{\boldsymbol{n}}_m) d\Omega_{dA \to dA_m}$$
$$- \sum_j \rho(\omega_m \to \omega_m, \widehat{\boldsymbol{n}}_m \to \widehat{\boldsymbol{n}}'_m) d\Omega_{dA \to dA_m},$$
(S.25)

where we have replaced $dA_m \cos\theta_m$ by $d\Omega_{dA \to dA_m}$ and split the summation over $m$. In the limit as $d\Omega_{dA \to dA_m} \to 0$, the summations over $m$ become hemispherical integrals:

$$e(\omega_m, \widehat{\boldsymbol{n}}'_m) - a(\omega_m, -\widehat{\boldsymbol{n}}'_m) = \int \rho(\omega_m \to \omega_m, -\widehat{\boldsymbol{n}}'_m \to -\widehat{\boldsymbol{n}}_m) d\Omega_{dA \to dA_m}$$
$$- \int \rho(\omega_m \to \omega_m, \widehat{\boldsymbol{n}}_m \to \widehat{\boldsymbol{n}}'_m) d\Omega_{dA \to dA_m},$$
(S.26)

where the integrations are over a hemisphere ($\int_0^{\pi/2} d\theta \int_0^{2\pi} d\phi$). Equation (S.26) is known result and can be found in prior work [11,13]. The hemispherical integrals properly account for scattering due to spatial modulation (in the form of a diffraction grating, surface roughness, etc.).

*No temporal modulation and no spatial modulation:* Now, if the graybody is not spatially modulated either, the BRDF is specular, meaning

$$e(\omega_m, -\widehat{\boldsymbol{n}}'_m) - a(\omega_m, \widehat{\boldsymbol{n}}'_m) = R(\omega_m, -\widehat{\boldsymbol{n}}_m) - R(\omega_m, \widehat{\boldsymbol{n}}_m),$$
(S.27)

where $R(\omega_m, \widehat{\boldsymbol{n}}_m)$ is the reflectance calculated assuming specular reflection, i.e., the amplitude squared of the Fresnel reflection coefficient. Equation (S.27) is a known result as well [10,12].

*No temporal modulation, no spatial modulation, and symmetric permittivity and permeability tensors:* If, further, the graybody has symmetric permittivity and permeability tensors, then $R(\omega_m, \widehat{\boldsymbol{n}}_m) = R(\omega_m, -\widehat{\boldsymbol{n}}_m)$. Thus Eq. (S.27) reduces to

$$e(\omega_m, \widehat{\boldsymbol{n}}'_m) = a(\omega_m, -\widehat{\boldsymbol{n}}'_m) \tag{S.28}$$

and we recover Kirchhoff's law of thermal radiation in its original form [17,21]. Without spatiotemporal modulation and with symmetric permittivity and permeability tensors, there is nothing to induce nonreciprocity, and the system is reciprocal.

*Other permutations of the constraints:* The most important constraint is (1) because temporal modulation causes both frequency conversion and scattering into multiple directional channels. Suppose the graybody is temporally modulated but not spatially modulated. This means $k_{zm} = \sqrt{(\omega_m/c)^2 - |\boldsymbol{k}_{\parallel,0}|^2}$, meaning each mode $m$ propagates in a different direction (even though longitudinal momentum is conserved). Therefore, even if we impose constraints (2) and (3), Eq. (S.24) still holds. On the other hand, (2) is the least important constraint because spatial modulation does not induce nonreciprocity. It can be used to tailor nonreciprocity induced by temporal modulation (Eq. (S.24)) or asymmetric permittivity and/or permeability tensors (Eq. (S.26)), but by itself, it does not violate Kirchhoff's law of thermal radiation.

**Table 1: Versions of Kirchhoff's law of thermal radiation**

| Constraints | | | Equation |
|---|---|---|---|
| $\bar{\bar{\varepsilon}} \neq \bar{\bar{\varepsilon}}(t)$ $\bar{\bar{\mu}} \neq \bar{\bar{\mu}}(t)$ | $\bar{\bar{\varepsilon}} \neq \bar{\bar{\varepsilon}}(\boldsymbol{r})$ $\bar{\bar{\mu}} \neq \bar{\bar{\mu}}(\boldsymbol{r})$ | $\bar{\bar{\varepsilon}} = \bar{\bar{\varepsilon}}^T$ $\bar{\bar{\mu}} = \bar{\bar{\mu}}^T$ | |
| | ✓ | | (S.24) |
| | | ✓ | |
| | ✓ | ✓ | |
| ✓ | | | (S.26) |
| ✓ | ✓ | | (S.27) |
| ✓ | | ✓ | (S.28) |
| ✓ | ✓ | ✓ | |

**Simplifying and interpreting Eq. (S.24):** In Eq. (S.24), if the modulation frequency $\Omega$ is small compared to the center frequency $\omega_0$, then it can be argued that $\omega_n \approx \omega_m$ and $I_b(\omega_n, T)/$

$I_b(\omega_m, T) \approx 1$. This may not be true if $n - m \gg \Omega/\omega_0$, but the diffraction efficiencies of higher order modes tend to be very low, so $\rho(\omega_m \to \omega_n, -\hat{\boldsymbol{n}}'_m \to -\hat{\boldsymbol{n}}_n)$ and $\rho(\omega_n \to \omega_m, \hat{\boldsymbol{n}}_n \to \hat{\boldsymbol{n}}'_m)$ would be close to zero anyway. This also implies that only the first few terms of the summation over $n$ in Eq. (S.24) matter:

$$e(\omega_m, \hat{\boldsymbol{n}}'_m) - a(\omega_m, -\hat{\boldsymbol{n}}'_m)$$

$$= [\rho(\omega_m \to \omega_m, -\hat{\boldsymbol{n}}'_m \to -\hat{\boldsymbol{n}}_m) - \rho(\omega_m \to \omega_m, \hat{\boldsymbol{n}}_m \to \hat{\boldsymbol{n}}'_m)]dA_m \cos\theta_m$$

$$+ [\rho(\omega_m \to \omega_{m+1}, -\hat{\boldsymbol{n}}'_m \to -\hat{\boldsymbol{n}}_{m+1}) - \rho(\omega_{m+1} \to \omega_m, \hat{\boldsymbol{n}}_{m+1} \to \hat{\boldsymbol{n}}'_m)]dA_{m+1} \cos\theta_{m+1}$$

$$+ [\rho(\omega_m \to \omega_{m-1}, -\hat{\boldsymbol{n}}'_m \to -\hat{\boldsymbol{n}}_{m-1}) - \rho(\omega_{m-1} \to \omega_m, \hat{\boldsymbol{n}}_{m-1} \to \hat{\boldsymbol{n}}'_m)]dA_{m-1} \cos\theta_{m-1}$$

$$+ \cdots \text{ (higher order terms)} \qquad (S.29)$$

In our experiments, $m = 0$ (and $\omega_0 = 30$ THz). Applying Eq. (S.29), we have that

$$e(\omega_0, \hat{\boldsymbol{n}}'_0) - a(\omega_0, -\hat{\boldsymbol{n}}'_0)$$

$$= [\rho(\omega_0 \to \omega_0, -\hat{\boldsymbol{n}}'_0 \to -\hat{\boldsymbol{n}}_0) - \rho(\omega_0 \to \omega_0, \hat{\boldsymbol{n}}_0 \to \hat{\boldsymbol{n}}'_0)]dA_0 \cos\theta_0$$

$$+ [\rho(\omega_0 \to \omega_{+1}, -\hat{\boldsymbol{n}}'_0 \to -\hat{\boldsymbol{n}}_{+1}) - \rho(\omega_{+1} \to \omega_0, \hat{\boldsymbol{n}}_{+1} \to \hat{\boldsymbol{n}}'_0)]dA_{+1} \cos\theta_{+1} \qquad (S.30)$$

$$+ [\rho(\omega_0 \to \omega_{-1}, -\hat{\boldsymbol{n}}'_0 \to -\hat{\boldsymbol{n}}_{-1}) - \rho(\omega_{-1} \to \omega_0, \hat{\boldsymbol{n}}_{-1} \to \hat{\boldsymbol{n}}'_0)]dA_{-1} \cos\theta_{-1}$$

$$+ \cdots \text{ (higher order terms)}.$$

For travelling-wave modulations such as the one used in our experiments, only the scattering from $\omega_0$ to $\omega_{\pm 1}$ is significant and specular reflection is reciprocal (independently confirmed by COMSOL simulations, shown in the main text Fig. 4 and described in Supplementary Note S4). In other words, $\rho(\omega_0 \to \omega_{\pm 1}, -\hat{\boldsymbol{n}}'_0 \to -\hat{\boldsymbol{n}}_{\pm 1}) \gg \rho(\omega_j \to \omega_i, -\hat{\boldsymbol{n}}'_j \to -\hat{\boldsymbol{n}}_i)$ ($j \neq 0, i \neq \pm 1$, and $i \neq j$) and $\rho(\omega_0 \to \omega_0, -\hat{\boldsymbol{n}}'_0 \to -\hat{\boldsymbol{n}}_0) = \rho(\omega_0 \to \omega_0, \hat{\boldsymbol{n}}_0 \to \hat{\boldsymbol{n}}'_0)$. Therefore, in the context of our experiments, Eq. (S.30) reduces to the relatively simple equation

$$e(\omega_0, \hat{\bm{n}}_0') - a(\omega_0, -\hat{\bm{n}}_0')$$
$$= \rho(\omega_0 \to \omega_{+1}, -\hat{\bm{n}}_0' \to -\hat{\bm{n}}_{+1}) + \rho(\omega_0 \to \omega_{-1}, -\hat{\bm{n}}_0' \to -\hat{\bm{n}}_{-1}) \tag{S.31}$$

Equation (S.31) states that in our experiments, the existence of a nonzero $\rho(\omega_0 \to \omega_{\pm 1}, -\hat{\bm{n}}_0' \to -\hat{\bm{n}}_{\pm 1})$ implies that $e(\omega_0, \hat{\bm{n}}_0') \neq a(\omega_0, -\hat{\bm{n}}_0')$, i.e., Kirchhoff's law of thermal radiation is violated).